%% file: main.tex
\documentclass{article}

\PassOptionsToPackage{numbers, compress}{natbib}
\usepackage[preprint]{neurips_2026}          % default = main track, anonymous
\usepackage[utf8]{inputenc}
\usepackage[T1]{fontenc}
\usepackage{hyperref}
\usepackage{url}
\usepackage{booktabs}
\usepackage{makecell}
\usepackage{multirow}
\usepackage{caption}
\usepackage{amsfonts}
\usepackage{amsmath}
\usepackage{amssymb}
\usepackage{nicefrac}
\usepackage{microtype}
\usepackage{threeparttable} % table footnotes below bottom rule
\usepackage{graphicx}
\usepackage{dsfont}
\usepackage{tabularx}       
\usepackage{enumitem}        % better enumerate control
\usepackage{subfig}
\usepackage[table]{xcolor}
\usepackage{bm}
\usepackage{pifont}          % provides \ding{51} (✓) and \ding{55} (✗)
\usepackage{algorithm}       % algorithm float
\usepackage{algpseudocode}   % \Procedure, \For, \State, etc.
\usepackage{pgfplots}        % inline bar charts
\pgfplotsset{compat=1.18}

\newcommand{\num}[1]{\scalebox{0.95}{$#1$}}
\newcommand{\best}[1]{\cellcolor{green!10}{\scalebox{0.95}{$\bm{#1}$}}}
\newcommand{\second}[1]{\scalebox{0.95}{$\underline{#1}$}}

\newcommand{\numnum}[1]{\scalebox{0.9}{$#1$}}
\newcommand{\bestbest}[1]{\cellcolor{green!10}{\scalebox{0.9}{$\bm{#1}$}}}
\newcommand{\secondsecond}[1]{\scalebox{0.9}{$\underline{#1}$}}

\graphicspath{{figures/}}

\title{Dynamic Deployment of Mobile Charging Trucks During Natural Disaster Evacuation: An Offline-to-Online Framework}

\author{%
  Rui Ma \\
  New York University\\
  \texttt{rm6968@nyu.edu} \\
  \And
  Zilin Bian \thanks{Corresponding author.}\\
  Rochester Institute of Technology\\
  \texttt{zilin.bian@rit.edu} \\
  \And
  Kaan Ozbay \\
  New York University\\
  \texttt{kaan.ozbay@nyu.edu} \\
}

\begin{document}

\maketitle
% ============================================================
%  Abstract
% ============================================================
\begin{abstract}
During large-scale evacuations, concentrated electric vehicle (EV) charging demand can overload fixed charging stations (FCSs), leading to prolonged waiting time and increased risk exposure. To address this challenge, this study proposes dynamically deploying mobile charging trucks (MCTs) to complement FCSs, and develops an Adaptive Risk-aware MCT Deployment (ARMD) framework for real-time operation. It divides the MCT deployment into two problems: risk-aware allocation of MCTs among FCSs and dynamic routing of MCTs to the assigned FCSs, and solves them under an offline-to-online paradigm. The resource allocation problem is formulated as a decentralized partially observable Markov decision process, and a multi-agent proximal policy optimization (MAPPO)-based policy is developed to coordinate multiple MCTs under decentralized observations. The policy is pre-trained offline in an evacuation simulator and adaptively refined online according to current evacuation context. For routing, a spatio-temporal travel time predictor is developed to support rolling-horizon route updates. The proposed framework is evaluated in a simulated hurricane evacuation environment built using real-world data from Hillsborough County, Florida. Experiments show that ARMD consistently outperforms offline optimization, online heuristic dispatch, and rolling-horizon optimization in reducing risk exposure. For demand perturbation scenarios, ARMD reduces average risk exposure by up to 71.1\%, relative to the baseline without MCTs. In the case of fixed e-vehicle charging infrastructure or road link failures, ARMD achieves 39.3\% to 60.5\% reduction in average risk exposure, with its advantages becoming more pronounced as the severity of disruption increases. These results demonstrate the effectiveness and robustness of ARMD in enhancing mobile charging operations for realistic scenarios of uncertain evacuation conditions.
\end{abstract}

% ============================================================
%  Paper body
% ============================================================
\input{introduction}
\input{literature_review}
\input{problme_formulation}
\input{methodology}
\input{experiments}
\input{implication}
\input{conclusion}

% ============================================================
 % Acknowledgments
% ============================================================
\begin{ack}
  This work was funded by the C2SMART transportation center led by New York University's Tandon School of Engineering.
\end{ack}

% ============================================================
%  References
% ============================================================
\bibliographystyle{unsrtnat}
\bibliography{references}

% ============================================================
%  Appendix
% ============================================================
\newpage

\input{appendix}
% ============================================================
%  NeurIPS Paper Checklist  (required — do NOT remove)
% ============================================================
% \newpage
% \input{checklist}

\end{document}

%% file: introduction.tex
\section{Introduction}
\label{sec:introduction}
Large-scale evacuation planning is a crucial component of preventive measures against natural disasters \cite{purba2022evacuation}. Taking hurricanes as an example, 27 evacuation orders were issued across the United States (US) between 2015 and 2025. In the meantime, as global efforts toward decarbonization accelerate, the electrification of road transportation is likely to reshape the landscape of emergency mobility. While electric vehicles (EVs) are essential instruments for sustainable mobility \cite{wang2025assessing, liu2025comprehensive}, their operational characteristics introduce possible vulnerabilities during disasters. Compared to internal combustion engine vehicles, EVs typically have shorter driving ranges \cite{USDOE2018}, depend on a relatively sparse network of charging infrastructure \cite{xmap2024, USDOE2025}, and require substantially longer dwelling times for energy replenishment \cite{USDOT2024}. Under routine traffic conditions, these constraints are manageable by optimizing charging schedules of individual vehicles \cite{elghanam2024optimization, zhao2024reinforcement}. 
However, during mass evacuations, these limitations are exacerbated by the temporal concentration of traffic flows, which can generate localized surges in charging demand. Consequently, static charging networks designed for routine steady-state demand may struggle to accommodate the transient yet intense loads generated during evacuation.

\begin{figure*}[!t]
\centering
\includegraphics[width=0.82\textwidth]{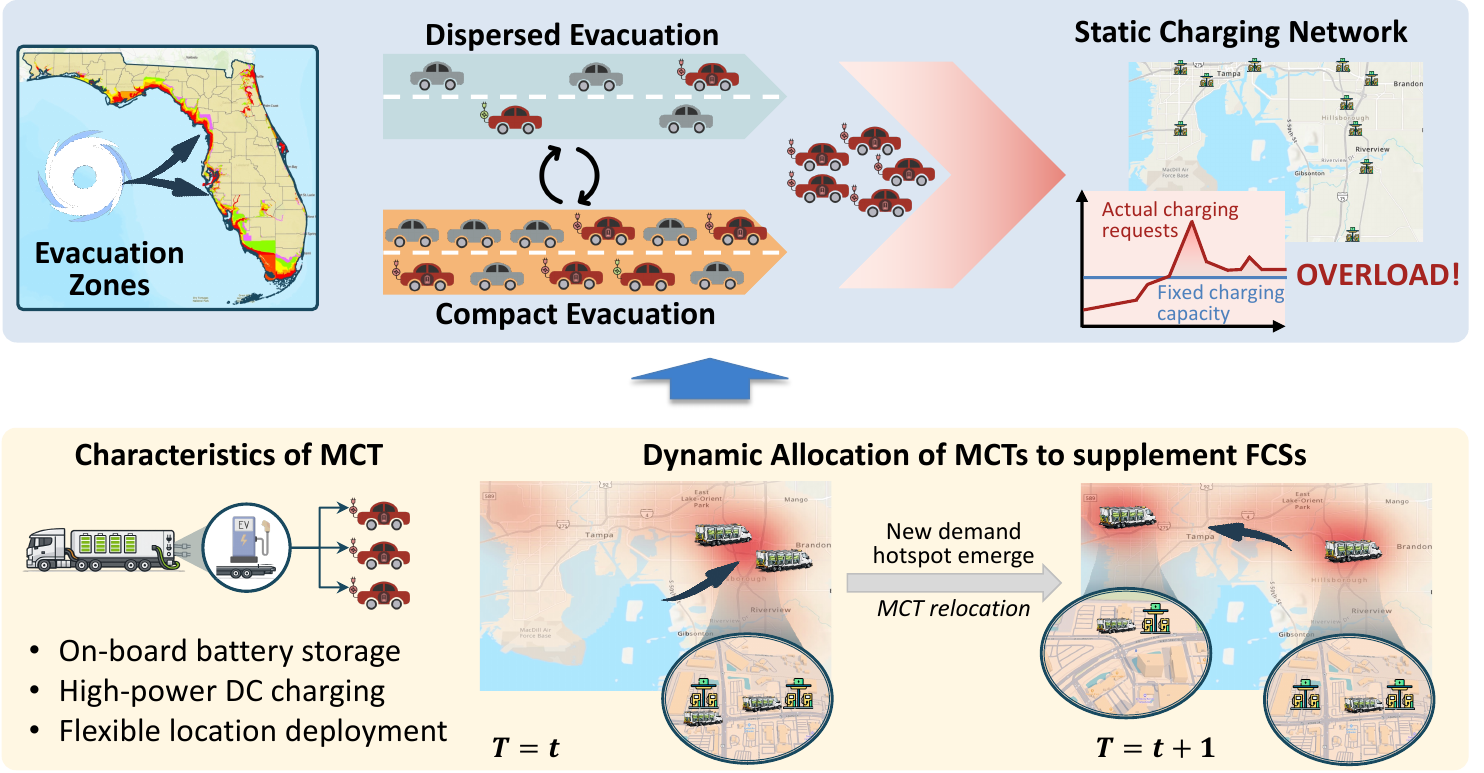}
\caption{Spatio-temporal demand–supply imbalance in charging network during evacuation, illustrated using hurricane evacuation as an example.}
\label{concept}
\end{figure*}

While prior studies have incorporated risk proximity considerations into long-term charging infrastructure planning to optimize the layout and capacity of fixed charging stations (FCSs) \cite{KCHAOUBOUJELBEN2021103376, ria, zhang2023optimal}, such static configurations lack the flexibility required under event-driven evacuation demand. To complement fixed charging networks, "mobile charging trucks" (MCTs) have emerged as a promising flexible supply of charging resources. Equipped with onboard energy storage and high-power chargers, MCTs can be strategically relocated to temporarily relieve overstressed stations \cite{jeon2021optimal}. In recent hurricanes, emergency agencies have pre-positioned transportable charging units along evacuation corridors to mitigate congestion at critical nodes \cite{FL_mobile}. Rather than expanding permanent infrastructure, dynamically allocating a fleet of MCTs to FCSs offers a responsive approach to reshaping charging supply during evacuation. 
As shown in Figure~\ref{concept}, this study aims to develop a decision-support framework for the real-time deployment of MCTs to coordinate with FCSs, thus improving the resilience and responsiveness of charging systems during evacuation. However, deploying MCTs under evacuation conditions presents several challenges:
\begin{itemize}
    \item \textbf{Adaptive deployment:} Evacuation traffic evolves continuously as evacuees move inland from coastal hazard zones to safer areas, generating uneven and time-varying charging demand across the network \cite{feng2020can}. 
    Static allocation rules or pre-calculated schedules exhibit limited adaptability to accommodate such dynamic demand patterns. Given the time-critical nature of emergency evacuation, the challenge lies in developing a decision mechanism that can adapt to stochastic demand evolution and generate deployment responses with minimal latency.
    \item \textbf{Decentralized decision-making: }As the number of FCSs and MCTs increases, the complexity of dispatch decisions grows exponentially, making global optimization across the large-scale network computationally intractable. Beyond these technical hurdles, evacuation operations are often inherently decentralized, as different authorities typically deploy emergency resources based on localized safety protocols and regional priorities, which constrains the feasibility of fully centralized dispatch. Therefore, the challenge lies in achieving effective fleet-level coordination while enabling decentralized decision-making. 

    \item \textbf{Anticipatory network response: }Since MCTs must physically traverse the network to provide relief to FCSs, their effectiveness is highly sensitive to the travel-time reliability and response delay. As evacuation flows and hazards' impacts continuously alter traffic conditions across the road network, relying on instantaneous travel time snapshots provides a static and potentially myopic view of the network. The challenge lies in anticipating travel-time dynamics and incorporating predictive information into dispatch decisions to facilitate timely intervention. 
\end{itemize}

This study proposes an Adaptive Risk-aware MCT Deployment (ARMD) framework to address the above challenges. The objective of ARMD is to deploy MCTs efficiently, thereby reducing the risk exposure of EV evacuees waiting for charging during evacuation. 
It divides the MCT deployment into two coupled decision problems: risk-aware allocation and dynamic routing, and follows an offline-to-online paradigm to support adaptive decision-making based on real-time system conditions. For MCT allocation, the problem is formulated as a decentralized partially observable Markov decision process (Dec-POMDP). A multi-agent proximal policy optimization (MAPPO)-based policy is developed to generate decentralized MCT allocation decisions through the actor network, while the critic network facilitates fleet-level coordination during training. To improve decision quality in decentralized settings with partial observations, a periodic observation augmentation mechanism is incorporated into MAPPO to enhance situational awareness without resorting to fully centralized control. During deployment, the pre-trained policy is refined by a scenario-specific online fine-tuning mechanism according to the current evacuation context. For MCT routing, a spatio-temporal travel time prediction module (STPM) is trained offline to anticipate short-term travel-time dynamics of evacuation traffic. After the target FCS of each MCT is determined by the allocation policy, the routing strategy is updated online in rolling-horizon based on STPM predictions. 
By integrating allocation and routing within the offline-to-online learning framework, the proposed ARMD aims to provide timely, scalable, and adaptive mobile charging support for evacuation charging systems.

We evaluated the effectiveness of the proposed ARMD in a simulation environment built using real-world data from Hillsborough County, Florida. The main contributions of this study are summarized as follows:
\begin{itemize}
    \item To the best of our knowledge, this study is the first to investigate dynamic, real-time MCT deployment for charging support under highly evolving evacuation conditions. It proposes ARMD, an offline-to-online learning framework to enable adaptive MCT deployment decisions to complement FCSs.
    \item To address the computational complexity and practical limitations of fully centralized control, the MCT allocation problem is formulated as a Dec-POMDP, and a multi-agent reinforcement learning approach is developed to enable coordinated yet decentralized decision-making under spatio-temporally varying charging demand.
    \item To improve MCTs' routing efficiency under dynamic evacuation traffic, a travel time prediction module is incorporated to anticipate short-term congestion evolution and support online routing, thereby reducing the likelihood that MCTs become stuck in congestion and improving the timeliness of charging support.
\end{itemize}

%% file: literature_review.tex
\section{Literature review}
\subsection{Challenges and Strategies for EV Evacuation}
While EVs offer significant environmental advantages in the global decarbonization efforts \cite{li2015hidden}, they are constrained by inherent limitations such as longer charging times and shorter driving ranges, compared to internal combustion engine vehicles. These limitations can be exacerbated by extreme weather associated with natural disasters. For example, extreme heat common in hurricane season may increase auxiliary power consumption and reduce driving range. 
A recent study reports that at $95^\circ$F, EVs experience an approximate $8.5\%$ reduction in driving range compared with moderate temperature conditions at $75^\circ$F \cite{AAA2026temperature}. 
In the context of large-scale evacuations, such limitations can force EVs to make multiple charging stops over long-distance trips, rendering EV evacuees heavily dependent on the fixed public charging infrastructure \cite{adderly2018electric}.

To address these charging requirements, extensive research has been conducted.
On the demand side, several studies focus on evacuee-oriented approaches that aim to design evacuation routes and ensure the effective management of EV mobility \cite{purba2022evacuation, li2022optimal, torkey2024emergency}. The primary objective is to satisfy charging demands while preventing the excessive aggregation of evacuees at specific stations, thereby minimizing the total evacuation time. Some literature emphasizes that public transportation remains a key element in emergency planning \cite{zhang2022multi}.
On the supply side, considering the strong dependency of EVs on power networks, researchers investigate strategies to enhance the resilience of charging infrastructure networks \cite{chong2023resilience, chen2023robust, zhang2023optimal, purba2024refueling}. Strategies in this domain focus on increasing the spatial coverage of charging facilities and mitigating infrastructure risks to ensure reliability for extreme conditions.

\subsection{Deployment of MCTs}
MCTs offer a flexible solution to deliver energy to EVs, and existing research has verified their benefits in complementing fixed charging infrastructure \cite{chen2024measuring, afshar2021mobile, zhang2020mobile, wang2021hybrid}.
However, the deployment of MCTs inevitably brings new challenges. At the planning level, the primary challenge lies in the optimal design of depot locations and fleet configurations\cite{tang2020online, wang2021hybrid}. 
At the operational level, MCTs are commonly treated as a door-to-door on-demand service, traveling directly to a specific user's location upon receiving a charging request. Under this paradigm, the challenge shifts to effectively dispatching MCTs to meet fluctuating charging demand. Several studies develop reactive dispatching mechanisms to route MCTs after charging requests are received \cite{cui2020multi, kabir2021joint, afshar2022mobile, qureshi2024dynamic, li2025electric}.
However, strategies that rely solely on current requests are prone to falling into a myopic trap \cite{tang2020online}.  
To overcome this limitation, recent studies leverage future demand prediction within look-ahead frameworks, allowing the system to anticipate demand dynamics and optimize MCT relocation proactively \cite{tang2020online, he2025coordinated}.
Furthermore, to specifically optimize the spatial distribution of idle resources, Liu et al. introduce physics-inspired force models to guide the distribution of idle MCTs before requests occur \cite{liu2022mobile, liu2023placement}.

In the real world, several industrial applications have been deployed. NIO launched the Power Mobile service, which functions as a mobile power bank delivering energy upon user request \cite{nio_power_mobile}; Tesla deploys mobile charging stations powered by Megapack in high-traffic areas during holidays \cite {tesla_megapack}. 
Existing studies and growing commercial implementations have demonstrated the value of MCTs in supporting EV charging, but most operational deployment models are developed for routine or daily charging-service contexts. However, large-scale evacuations are characterized by rapidly evolving charging demand, time-varying traffic conditions, and possible infrastructure failures, which require MCTs to be dynamically dispatched in real time. To the best of our knowledge, dynamic MCT deployment for evacuation charging support under such evolving conditions remains largely unexplored.

\subsection{Dispatching and Scheduling Methods for MCTs}
From the methodological perspective, the dispatching and scheduling of MCTs have been formulated as integer programs (IP) or mixed-integer programs (MIP) \cite{ruaboacua2020optimization, kabir2021joint, jeon2021optimal}. Due to the combinatorial nature of vehicle routing and charging coordination, these formulations are typically NP-hard and become computationally intractable for large-scale systems. To improve computational efficiency, heuristic and metaheuristic algorithms are adopted to obtain near-optimal solutions within reasonable computational time, such as ant colony optimization algorithm \cite{moghaddam2021dispatch}, modified Clarke and Wright’s savings heuristic and genetic algorithm \cite{qureshi2022scheduling}. In addition, Ala et al. combine Q-learning with differential evolution algorithm to improve solution robustness \cite{ala2024dynamic}. To account for the charging demand uncertainty, Tang et al. model the MCT dispatching as a stochastic optimization problem and propose a scenario-sampling-based policy to enable real-time operation \cite{tang2020online}. Based on the lookahead rolling horizon value function approximation method, He et al. establish a two-stage scheduling framework, where value function approximations are trained offline from historical data while online rolling optimization updates decisions in response to real-time demand \cite{he2025coordinated}.

More recently, learning-based approaches have begun to emerge in this domain. Liu et al. employ a stacked long short-term memory (LSTM) model to predict future charging positions for idle MCTs and federated learning is adopted to accelerate model training \cite{liu2022mobile}. 
In addition, Qureshi et al. explore a distributed model-free reinforcement learning (RL) approach based on deep Q-networks to dynamically allocate MCTs in response to evolving charging requests \cite{qureshi2024dynamic}. The result demonstrates that RL-based methods can effectively handle the stochastic and real-time nature of MCT scheduling without requiring explicit system modeling.

%% file: problme_formulation.tex
\section{Problem Formulation}
Instead of optimizing static MCT placement, this study formulates MCT deployment as a dynamic, risk-aware decision-making problem under evolving evacuation conditions. The proposed formulation does not focus on traditional objectives such as minimizing travel time or operating cost, but explicitly aims to minimize queue-induced risk exposure experienced by EV evacuees waiting at FCSs. In this section, we provide the necessary definitions and formulates the dynamic MCT deployment problem.

Let $G=(\mathcal{N},\mathcal{E})$ denote the transportation network, where $\mathcal{N}$ denotes the set of nodes and $\mathcal{E}$ denotes the set of directed road links. A subset of nodes $\mathcal{F}\subseteq \mathcal{N}$ corresponds to FCSs. Each station $i\in\mathcal{F}$ is equipped with $m_i(t)$ available fixed chargers at time step $t$.
Here, the time-varying notation is used to capture stochastic charger failures during evacuation.

During evacuations, evacuees travel from hazard-prone areas to safe destinations. Among them, EV evacuees may generate en-route charging demand at different FCSs as their batteries deplete, resulting in time-varying queue lengths $Q_i(t)$ at station $i$.
Prolonged waiting for charging service implies prolonged risk exposure. Let $H(t)$ denote the global hazard at time $t$. The corresponding per-capita risk exposure at station $i$ is defined as $R_i(t)=\phi_i\!\bigl(H(t)\bigr)$, 
where $\phi_i(\cdot)$ maps the global hazard to local per-capita risk exposure experienced by evacuees waiting at station $i$. The aggregate risk exposure at station $i$ at time $t$ is then given by
\begin{equation}
\mathcal{R}_i(t)=R_i(t)\,Q_i(t).
\label{eq:station_risk_new}
\end{equation}

To supplement the fixed charging infrastructure, a fleet of MCTs, indexed by $k\in\mathcal{K}=\{1,\dots,K\}$, is deployed. Each MCT carries $m_k$ mobile chargers and can provide temporary charging support at selected FCSs. The evacuation horizon is discretized into time steps $\mathcal{T}=\{1,\dots,T\}$. Since an MCT must travel to its assigned FCS and remain there long enough to provide meaningful service, allocation decisions are updated only at a subset of decision epochs, denoted by $\mathcal{T}^{\mathrm{dec}}=\{\tau_1,\tau_2,\dots,\tau_M\}\subseteq \mathcal{T}$. The objective is to coordinate MCT deployment so as to relieve overloaded FCSs and minimize cumulative queue-induced risk exposure over the evacuation horizon:
\begin{equation}
\min \mathcal{R}^{\mathrm{tot}}
=
\sum_{t\in\mathcal{T}}\sum_{i\in\mathcal{F}}\mathcal{R}_i(t)
=
\sum_{t\in\mathcal{T}}\sum_{i\in\mathcal{F}}R_i(t)\,Q_i(t).
\label{eq:total_risk_new}
\end{equation}

Achieving Eq.~\eqref{eq:total_risk_new} requires solving two coupled problems:
\begin{itemize}
    \item \textbf{P1: MCT allocation.} At each decision epoch $\tau_m\in\mathcal{T}^{\mathrm{dec}}$, determine the target FCS for each MCT according to the current network condition.
    \item \textbf{P2: MCT routing.} Given the assigned target FCS, determine the route of each MCT to improve response efficiency under time-varying traffic conditions.
\end{itemize}

%% file: methodology.tex
\begin{figure*}[!t]
\centering
\includegraphics[width=0.9\textwidth]{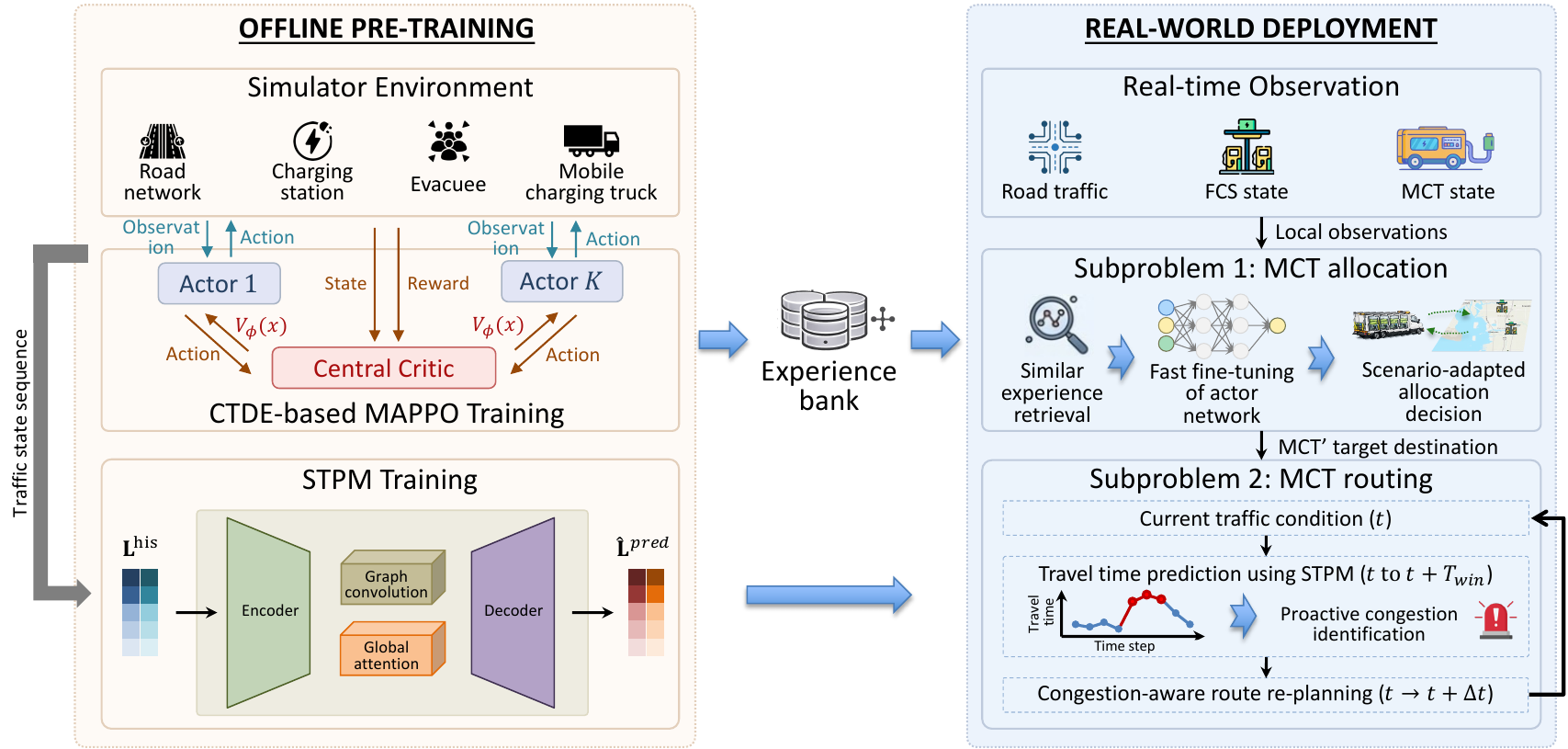}
\caption{Architecture of the ARMD framework. It divides MCT deployment into two coupled subproblems: allocation, which determines the target FCSs, and routing, which determines MCTs' travel paths. Both subproblems follow an offline-to-online paradigm, where policy and predictor are first trained offline in the evacuation simulator and then adaptively updated and applied during real-world deployment based on real-time charging and traffic states.}
\label{framework}
\end{figure*}

\section{Methodology}
\subsection{An Overview of the ARMD framework}
The ARMD framework is designed to adaptively utilize MCTs to mitigate the risk exposure experienced by EV evacuees while waiting for charging at FCSs. It focuses on two coupled subproblems: selecting the target FCSs to be served and determining the relocation routes of MCTs under evolving evacuation traffic, as illustrated in Figure~\ref{framework}.
Unlike traditional offline methods for resource allocation \cite{ozbay2013probabilistic}, ARMD combines offline learning with online updating to enable adaptive MCT deployment decisions in response to real-time and stochastic evacuation conditions.
In the offline phase, a simulator is developed to reproduce the evacuation environment, including traffic conditions, charging demand evolutions, FCS queue dynamics, and MCT operations. The simulator also incorporates stochastic infrastructure failures, so that the learned policy can be exposed to various conditions before deployment. Based on the simulations, a MAPPO-based policy is pre-trained to support risk-aware FCS selection, and an STPM is trained to anticipate short-term traffic dynamics.
In real-world deployment, ARMD operates as a closed-loop online decision process that continuously incorporates observations from the real-time system. FCS states reported by operators, including waiting charging demand and available chargers, are used to retrieve similar experiences from pre-training and fine-tune the allocation policy online. The refined policy then determines target FCS for each MCT according to evolving queue-induced risk. After target FCSs are assigned, the routing module starts to determine MCT relocation paths. Recent traffic observations collected from probe vehicles are fed into the trained STPM to generate short-term travel time predictions. Road disruption information, including road or bridge closures issued by emergency management agencies and evacuee-reported road failures, is incorporated to update network accessibility. Based on the predicted traffic dynamics and updated network conditions, MCT routes are updated online in rolling-horizon. This online updating mechanism enables ARMD to continuously adapt allocation and routing decisions to evolving environments.
The intermediate variables and parameters are summarized in Table~\ref{tab:notation}.

\begin{table}[htbp]
\centering
\caption{Summary of frequently used notations.}
\label{tab:notation}
\small
\begin{tabular}{>{\raggedright\arraybackslash}p{2.2cm} >{\raggedright\arraybackslash}p{5.8cm}}
\toprule
Symbol & Description \\
\midrule
\multicolumn{2}{l}{\textit{Sets and indices}} \\
\cmidrule(lr){1-2}
$\mathcal{N}, \mathcal{E}$ & Node set and road-link set \\
$\mathcal{F}, \mathcal{K}$ & FCS set and MCT set \\
$\mathcal{T}, \mathcal{T}^{\mathrm{dec}}$ & Discrete time steps and MCT allocation decision epochs \\
\midrule
\multicolumn{2}{l}{\textit{System and MCT states}} \\
\cmidrule(lr){1-2}
$H(t)$ & Global hazard at time $t$ \\
$Q_i(t), m_i(t), c_i(t)$ & Queue length, available fixed chargers, and MCTs serving FCS $i$ at time $t$ \\
$R_i(t)$ & Per-capita risk exposure at FCS $i$ at time $t$ \\
$u_k(t), m_k$ & Remaining charging capability and number of mobile chargers of MCT $k$ \\
$L_{k,i}(t)$ & Travel time from MCT $k$ to FCS $i$ at time $t$ \\
\midrule
\multicolumn{2}{l}{\textit{Dec-POMDP variables}} \\
\cmidrule(lr){1-2}
$x^m, o_k^m$ & Global state and local observation \\%  at decision epoch $\tau_m$ \\
$a_k^m, \mathbf{a}^m$ & Agent action and joint action \\%  at decision epoch $\tau_m$ \\
$r^m $ & Shared reward for agents \\ % over $[\tau_m,\tau_{m+1})$ \\
$\gamma$ & Discount factor \\
\bottomrule
\end{tabular}
\end{table}

\subsection{Problem I: MCT Allocation}
\subsubsection{Dec-POMDP Formulation}
To solve \textbf{P1}, the MCT allocation problem is formulated as a Dec-POMDP, defined by a tuple $\mathcal{G}
=
\left\langle
\mathcal{K},\mathcal{X},\{\mathcal{O}_k\}_{k\in\mathcal{K}},\{\mathcal{U}_k\}_{k\in\mathcal{K}},P,r,\gamma
\right\rangle$, 
where $\mathcal{K}$ is the agent set, $\mathcal{X}$ is the global state space, $\mathcal{O}_k$ and $\mathcal{U}_k$ are the observation and action spaces of agent $k$, respectively, $P$ is the state transition function, $r$ is the shared reward, and $\gamma\in(0,1)$ is the discount factor.
This decentralized formulation is adopted for two reasons. First, emergency resource deployment is typically carried out in a distributed manner, where local decisions are made based on regional information and operational priorities rather than by a fully centralized controller. Second, as the numbers of FCSs and MCTs increase, the joint action space of centralized dispatch grows combinatorially, making fully centralized real-time optimization increasingly impractical. The detailed descriptions are as follows:

\textbf{Agent.}
Each MCT is modeled as an agent. Thus, the agent set is $\mathcal{K}=\{1,2,\dots,K\}$. At each decision epoch $\tau_m$, every agent selects one target FCS to support.

\textbf{State. }
Let $x^m$ denote the global state observed at decision epoch $\tau_m$. It is defined as
\begin{subequations}
\label{eq:all_states} 
\begin{equation}
x^m=\bigl[H(\tau_m),\,S(\tau_m),\,M(\tau_m)\bigr].
\label{eq:global_state_new}
\end{equation}
where $H(\tau_m)$ is the global hazard, $S(\tau_m)$ summarizes current FCS conditions, and $M(\tau_m)$ summarizes MCT conditions.

The FCS state component is given by
\begin{equation}
S(\tau_m)=\left[Q_i(\tau_m),\,R_i(\tau_m),\,m_i(\tau_m),\,c_i(\tau_m)\right]_{i\in\mathcal{F}},
\label{eq:fcs_state_new}
\end{equation}
where $Q_i(\tau_m)$, $R_i(\tau_m)$, $m_i(\tau_m)$, and $c_i(\tau_m)$ denote the queue length, per-capita risk exposure, available fixed chargers, and the number of MCTs currently serving FCS $i$, respectively.

The MCT state component is given by
\begin{equation}
M(\tau_m)=\left[\left(u_k(\tau_m),\,\{L_{k,i}(\tau_m)\}_{i\in\mathcal{F}}\right)\right]_{k\in\mathcal{K}},
\label{eq:mct_state_new}
\end{equation}
\end{subequations}
where $u_k(\tau_m)$ denotes the remaining charging capability of MCT $k$, and $L_{k,i}(\tau_m)$ denotes the travel time from MCT $k$ to FCS $i$.

\textbf{Local observation.} Under decentralized execution, each MCT observes only a local subset of the global state. The local observation of agent $k$ at decision epoch $\tau_m$ is
\begin{subequations}
\begin{equation}
o_k^m=
\left[
H(\tau_m),\,
S_k^{\mathrm{obs}}(\tau_m),\,
M_k^{\mathrm{obs}}(\tau_m)
\right],
\label{eq:local_obs_new}
\end{equation}
where
\begin{small} 
\begin{equation}
S_k^{\mathrm{obs}}(\tau_m)=
\left[
Q_i(\tau_m), R_i(\tau_m),m_i(\tau_m),c_i(\tau_m)
\right]_{i\in\mathcal{F}_k^{\mathrm{obs}}(\tau_m)},
\end{equation}
\end{small}
\begin{equation}
M_k^{\mathrm{obs}}(\tau_m)=
\left(
u_k(\tau_m),\,
\{L_{k,i}(\tau_m)\}_{i\in\mathcal{F}_k^{\mathrm{obs}}(\tau_m)}
\right).
\end{equation}
\end{subequations}
Here, $\mathcal{F}_k^{\mathrm{obs}}(\tau_m)\subseteq\mathcal{F}$ denotes the subset of observable FCSs for agent $k$.

\textbf{Action.} At decision epoch $\tau_m$, agent $k$ selects one target FCS from its observable station set $a_k^m \in \mathcal{U}_k(o_k^m)=\mathcal{F}_k^{\mathrm{obs}}(\tau_m)$.
The corresponding joint action is denoted by $\mathbf{a}^m=(a_1^m,\dots,a_K^m)$.

\textbf{State transition.}
Given the current state $x^m$ and joint action $\mathbf{a}^m$, the evacuation environment evolves over the interval $[\tau_m,\tau_{m+1})$ according to evacuees' charging demand, MCT movement, charging service dynamics, and hazard evolution, and produces the next decision-epoch state $x^{m+1}$.

\textbf{Reward.}
To align the RL objective with the system objective in Eq.~\eqref{eq:total_risk_new}, we define the shared reward at each decision epoch as the negative cumulative queue-induced risk exposure over the decision interval. Maximizing the expected discounted return is therefore equivalent to minimizing cumulative queue-induced risk exposure over the evacuation horizon.
\begin{equation}
r^m=
-\sum_{t=\tau_m}^{\tau_{m+1}-1}\sum_{i\in\mathcal{F}}\mathcal{R}_i(t).
\label{eq:reward_new}
\end{equation}

\subsubsection{Offline Policy Pre-training}
Real-world deployment data for MCT allocation under large-scale evacuations are extremely limited, making direct policy learning from historical field data infeasible. To address this issue, we develop an evacuation simulator based on the underlying charging-service dynamics, MCT operations, and risk-evolution mechanism. The allocation policy is pre-trained offline through repeated interactions between the MCT agents and the simulator.

To learn a cooperative policy, we adopt MAPPO under the centralized training and decentralized execution (CTDE) paradigm \cite{yu2022surprising}. Since all MCTs are operationally homogeneous, a shared actor policy $\pi_{\theta}$ is used across agents, while a centralized critic $V_{\phi}(x^m)$ exploits the global state at decision epoch $\tau_m$ to provide value estimation. Under this design, agent $k$ selects its action according to $a_k^m \sim \pi_{\theta}(\cdot \mid o_k^m)$.

\textbf{Relational Station Encoding.}
\label{sec:encoder_PPO}
The allocation decision of an MCT depends not only on the status of an individual candidate FCS, but also on its relative priority with respect to other observable stations competing for limited mobile charging support. To capture inter-station dependency, we embed a relational station encoder into actor and critic networks.

For agent $k$ at decision epoch $\tau_m$, let $\mathbf{e}_{k,i}^m$
denote the embedded feature of observable station $i \in \mathcal{F}_k^{\mathrm{obs}}(\tau_m)$. 
Stacking all observable station embeddings yields $\mathbf{E}_k^m \in \mathbb{R}^{|\mathcal{F}_k^{\mathrm{obs}}(\tau_m)|\times d_{\mathrm{model}}}$.
A multi-head attention block is then applied to produce context-aware station representations:
\begin{equation}
\begin{aligned}
\mathbf{Q}_k^m &= \mathbf{E}_k^m\mathbf{W}_Q,\qquad
\mathbf{K}_k^m = \mathbf{E}_k^m\mathbf{W}_K,\qquad
\mathbf{V}_k^m = \mathbf{E}_k^m\mathbf{W}_V, \\
\widetilde{\mathbf{E}}_k^m
&=
\mathrm{MHA}\!\left(\mathbf{Q}_k^m,\mathbf{K}_k^m,\mathbf{V}_k^m\right).
\end{aligned}
\label{eq:relational_encoder}
\end{equation}
where $\mathbf{W}_Q$, $\mathbf{W}_K$, and $\mathbf{W}_V$ are trainable projection matrices, $\mathrm{MHA}(\cdot)$ denotes the multi-head attention operator.
Through this operation, the representation of each candidate station is refined by aggregating information from other observable stations, allowing the actor to compare the relative value of candidate FCSs based on cross-station relationships. The critic adopts the same relational encoder architecture, but takes the global state as input to estimate the value function.

\textbf{Periodic Observation Augmentation.}
A key challenge of decentralized decision making is insufficient situational awareness. The low observation range of agents can create an information bottleneck that will lead to poor coordination quality \cite{shao2023complementary}. Under evacuation conditions, charging demand can shift rapidly across regions. If each agent relies only on a strictly local observation range, emerging high-risk stations outside its local field of view may not be detected in time, which weakens cross-regional coordination.

To mitigate this issue without introducing persistent all-to-all communication, we incorporate a periodic observation augmentation mechanism. Let $\mathcal{F}_{k,\mathrm{loc}}^m$ denote the local observable station set of agent $k$ at decision epoch $\tau_m$. Every $\Delta_{\mathrm{aug}}$ decision epochs, the local observation is augmented with the full station set:
\begin{equation}
\mathcal{F}_k^{\mathrm{obs}}(\tau_m)=
\begin{cases}
\mathcal{F}, & m \bmod \Delta_{\mathrm{aug}} = 0,\\[4pt]
\mathcal{F}_{k,\mathrm{loc}}^m, & \text{otherwise}.
\end{cases}
\label{eq:obs_set_aug}
\end{equation}

In this way, each agent operates under decentralized partial observation most of the time, while periodically receiving a global snapshot of the charging network. This mechanism preserves the scalability and practicality of decentralized execution, while still enabling agents to incorporate broader network information for improved cross-regional coordination under rapidly evolving evacuation conditions.

\subsubsection{Online Policy Fine-tuning}
The pre-trained policy provides an initialization for deployment, but the environment distribution encountered online may deviate from that seen during offline training. Direct online adaptation using newly collected real-time trajectories is highly sample inefficient in high-dimensional settings \cite{liu2025active}, especially under emergency evacuations where online interaction is extremely limited. To improve policy adaptivity while preserving update stability, we introduce a retrieval-augmented online fine-tuning mechanism.

After obtaining the pre-trained actor $\pi_{\theta}^{\mathrm{pre}}$, we roll it out offline in the simulator under diverse set of stochastic evacuation scenarios to construct an experience bank $\mathcal{B} = \{ (\mathbf{c}_k^{\,n}, o_k^{\,n}, a_k^{\,n}, r^{\,n}, \hat{A}^{\,n}) \}_{n=1}^{N}$, 
where $\mathbf{c}_k^{\,n}$ denotes the local context descriptor associated with agent $k$ in sample $n$, and $\hat{A}^{\,n}$ is the corresponding advantage estimate. At decision epoch $\tau_m$, the local context descriptor observed by agent $k$ is defined as
\begin{subequations}
\begin{equation}
\mathbf{c}_k^m
=
\left[
H(\tau_m),\;
\frac{Q_{k,\mathrm{tot}}^m}{m_{k,\mathrm{tot}}^m}\;
\right],
\label{eq:local_context_descriptor}
\end{equation}
where
\begin{equation}
\begin{aligned}
Q_{k,\mathrm{tot}}^m
&=
\sum_{i\in\mathcal{F}_k^{\mathrm{obs}}(\tau_m)} Q_i(\tau_m), \\
m_{k,\mathrm{tot}}^m
&=
\sum_{i\in\mathcal{F}_k^{\mathrm{obs}}(\tau_m)} m_i(\tau_m).
\end{aligned}
\label{eq:local_context_stats}
\end{equation}
\end{subequations}
Here, $Q_{k,\mathrm{tot}}^m$ and $m_{k,\mathrm{tot}}^m$ summarize the observable charging load and available fixed charging capacity perceived by agent $k$. During deployment, instead of using all samples in $\mathcal{B}$ uniformly, agent $k$ retrieves the $K_{\mathrm{ret}}$ most similar samples $\mathcal{N}_k^m$ according to its current local context descriptor using the Euclidean similarity.

\textbf{Advantage-Weighted Actor Adaptation.}
Using the retrieved subset $\mathcal{N}_k^m$, a lightweight actor-only adaptation is then applied. Each retrieved sample is associated with a pre-computed advantage value $\hat{A}^{\,n}$ stored in the experience bank, which is used to assign larger importance to more valuable behaviors. Specifically, the sample weight is defined as 
\begin{subequations}
\begin{equation}
    w_n
    =
    \exp\!\left(
    \frac{\hat{A}^{\,n}}{\beta}
    \right),
    \label{eq:adv_weight}
\end{equation}
where $\beta>0$ is a temperature parameter controlling the sharpness of the weighting scheme.

Starting from the pre-trained actor $\pi_{\theta}^{\mathrm{pre}}$, we refine the actor by maximizing the weighted log-likelihood of retrieved actions while regularizing the deviation from the pre-trained policy:
\begin{equation}
\begin{aligned}
\max_{\theta}\quad
& \sum_{n\in\mathcal{N}_k^m}
w_n\,
\log \pi_{\theta}(a_k^{\,n}\mid o_k^{\,n}) \\
\text{s.t.}\quad
& \frac{1}{|\mathcal{N}_k^m|}
\sum_{n\in\mathcal{N}_k^m}
D_{\mathrm{KL}}\!\left(
\pi_{\theta}(\cdot\mid o_k^{\,n})
\,\|\, 
\pi_{\theta}^{\mathrm{pre}}(\cdot\mid o_k^{\,n})
\right)
\le \varepsilon_{\mathrm{KL}} .
\end{aligned}
\label{eq:online_aw_constrained}
\end{equation}
\end{subequations}
where $\lambda_{\mathrm{KL}}>0$ controls the strength of the trust-region regularization. Only the actor is adapted online, while the centralized critic remains fixed during deployment. At each decision epoch, the actor is re-initialized to $\pi_{\theta}^{\mathrm{pre}}$ and then updated using only the retrieved subset $\mathcal{N}_k^m$. This epoch-wise reset prevents policy drift from accumulating across deployment steps and ensures conservative adaptation to the current local operating condition.

\subsection{Problem II: MCT Routing}
\subsubsection{Online Path Planning}
Once assigning MCT $k$ to a target charging station $a_k^m \in \mathcal{F}$ at decision epoch $\tau_m$, the corresponding route should not be fixed, because link travel times may vary substantially during evacuation due to congestion propagation, lane blockages, and spatially uneven demand. Therefore, to solve \textbf{P2} we formulate MCT routing as an online path planning problem under time-varying traffic conditions.

Let $\mathbf{L}(t)\in\mathbb{R}^{|\mathcal{E}|}$ denote the edge-level travel time vector at time $t$, where each component corresponds to one link $e\in\mathcal{E}$. Suppose that at time $t$, MCT $k$ is located at node $n_k(t)$ and is traveling toward its assigned station $a_k^m$. Instead of determining the route solely from the current travel time snapshot $\mathbf{L}(t)$, we adopt a prediction-based rolling-horizon routing strategy. Specifically, at each routing update instant $t$, the travel time prediction module takes the most recent history
\[
\mathbf{L}^{\mathrm{his}}(t)
=
\left[
\mathbf{L}(t-T_{\mathrm{win}}+1),\,
\dots,\,
\mathbf{L}(t)
\right]
\in
\mathbb{R}^{T_{\mathrm{win}}\times |\mathcal{E}|},
\]
as input and predicts the future edge-level travel times over the next $T_{\mathrm{win}}$ periods:
\[
\widehat{\mathbf{L}}^{\mathrm{pred}}(t)
=
\left[
\widehat{\mathbf{L}}(t+1),\,
\dots,\,
\widehat{\mathbf{L}}(t+T_{\mathrm{win}})
\right]
\in
\mathbb{R}^{T_{\mathrm{win}}\times |\mathcal{E}|}.
\]

Based on these forecasts, the routing module computes a path from the current location $n_k(t)$ to the assigned station $a_k^m$. Let $\mathcal{P}\!\left(n_k(t),a_k^m\right)$ denote the set of feasible paths between them. The routing decision is given by
\begin{equation}
p_{k,t}^{*}
=
\arg\min_{p\in\mathcal{P}(n_k(t),a_k^m)}
\widehat{C}(p\,|\,t),
\label{eq:routing_problem}
\end{equation}
where $\widehat{C}(p\,|\,t)$ denotes the predicted travel cost of path $p$ under the forecast travel times. After obtaining $p_{k,t}^{*}$, MCT $k$ follows this route for one routing update interval $\Delta t$. Once new traffic information becomes available at $t+\Delta t$, the travel time predictor is invoked again and the remaining route is recomputed from the updated location. This process is repeated until the MCT reaches its assigned charging station. Through this rolling-horizon procedure, route planning is continuously refined online according to predicted traffic evolution, enabling the MCT to respond adaptively to changing network conditions and proactively avoid impending congestion.

\subsubsection{Spatio-Temporal Travel Time Prediction Model}
To support routing in Eq.~\eqref{eq:routing_problem}, we develop STPM, which is trained offline and deployed online to provide short-term travel time forecasts for online MCT routing. Building such a predictor involves two main challenges. First, congestion in evacuation traffic often propagates locally across adjacent road segments, particularly near bottlenecks and access roads to FCSs, making it necessary to capture localized spatial dependencies. Second, evacuation traffic also exhibits strong network-wide interactions. For example, road closures or link failures may trigger large-scale rerouting as evacuees avoid disrupted areas, thereby inducing nonlocal effects that reshape traffic states across the network. To address these challenges, STPM combines a dynamic graph convolution to capture localized topology-aware dependencies with a global multi-head attention block to model long-range spatio-temporal interactions.

\textbf{Dynamic Graph Convolution.}
To capture the localized interactions, we apply graph convolution on the edge-based road network representation \cite{kipf2016semi}. Let $\mathbf{H}^{(l)}\in\mathbb{R}^{|\mathcal{E}|\times d_{\mathrm{model}}}$ denote the hidden representation at layer $l$, $\widetilde{\mathbf{A}}_{\mathrm{edge}}$ and $\widetilde{\mathbf{D}}_{\mathrm{edge}}$ denote the edge-adjacency matrix and degree matrix of the road network. 
Using only a static adjacency structure $\widetilde{\mathbf{A}}_{\mathrm{edge}}$ is insufficient under evacuation, because the correlations between neighboring links are likely to change over time. That is, the extent to which one link affects the travel time evolution of its adjacent links is state-dependent and may vary as evacuation traffic evolves.
We therefore introduce a dynamic spatial weighting matrix \cite{li2025digital}
\begin{subequations}
\begin{equation}
\mathbf{S}^{(l)}
=
\mathrm{softmax}
\left(
\frac{\mathbf{H}^{(l)}\mathbf{H}^{(l)\top}}
{\sqrt{d_{\mathrm{model}}}}
\right)
\in
\mathbb{R}^{|\mathcal{E}|\times|\mathcal{E}|},
\label{eq:dynamic_spatial_weight}
\end{equation}
which adaptively modulates the local aggregation strength according to the current hidden representation. The dynamic graph convolution update is then defined as
\begin{equation}
\mathbf{U}_{\mathrm{spat}}^{(l)}
=
\sigma\!\left(
\left(
\widetilde{\mathbf{D}}_{\mathrm{edge}}^{-1}
\widetilde{\mathbf{A}}_{\mathrm{edge}}
\right)
\odot
\mathbf{S}^{(l)}
\right.
\mathbf{H}^{(l)}\mathbf{W}^{(l)}
\left.
\right),
\label{eq:dynamic_gcn}
\end{equation}
\end{subequations}
where $\mathbf{W}^{(l)}$ is a trainable weight matrix and $\sigma(\cdot)$ is the nonlinear activation function.

\textbf{Global Attention.}
To capture the nonlocal temporal interactions, STPM introduces a temporal-aware multi-head global attention \cite{li2024multi}.
Let $\mathbf{Z}^{(l)}$ denote the input hidden representation to the temporal attention at layer $l$. Temporal convolution is first applied when constructing the query and key representations, so that each time step can incorporate short-range temporal context from its neighboring periods. The temporal-aware multi-head global attention is defined as
\begin{subequations}
\begin{equation}
\mathbf{U}_{\mathrm{temp}}^{(l)}
=
\mathrm{Concat}\!\left(
\mathrm{head}_1^{(l)},\dots,\mathrm{head}_{H_{\mathrm{att}}}^{(l)}
\right)\mathbf{W}^{O},
\label{eq:temp_mha}
\end{equation}

\begin{equation}
\mathrm{head}_h^{(l)}
=
\mathrm{Attention}\!\left(
\Phi_h^{Q}\mathbf{Z}^{(l)},
\Phi_h^{K}\mathbf{Z}^{(l)},
\mathbf{Z}^{(l)}\mathbf{W}_h^{V}
\right).
\label{eq:temp_head}
\end{equation}
\end{subequations}
Here, $\Phi_h^{Q}$ and $\Phi_h^{K}$ denote the learnable temporal convolution kernels for forming the query and key representations, $\mathbf{W}_h^{V}$ and $\mathbf{W}^{O}$ are trainable projection matrices, and $H_{\mathrm{att}}$ is the number of attention heads. 
Under this design, temporal convolution first extracts local temporal trends from neighboring periods, while the global attention mechanism allows each position in the traffic state sequence to attend to every other position within the input window.

%% file: experiments.tex
\section{Experiments}
\subsection{Simulator Setup}
\label{sim_setup}

\begin{figure*}[!t]
\centering
\subfloat[]{
    \includegraphics[width=0.32\textwidth]{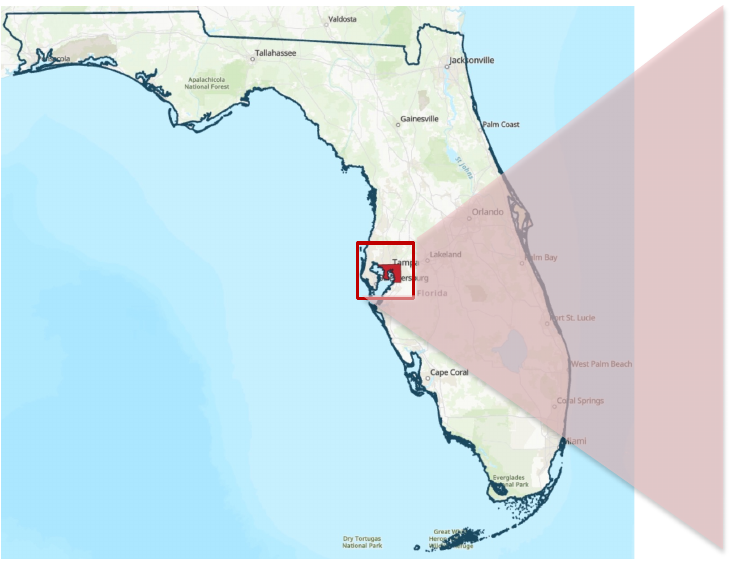}
    \label{fig:a}
}%
\hspace{-4mm}%
\subfloat[]{
    \includegraphics[width=0.31\textwidth]{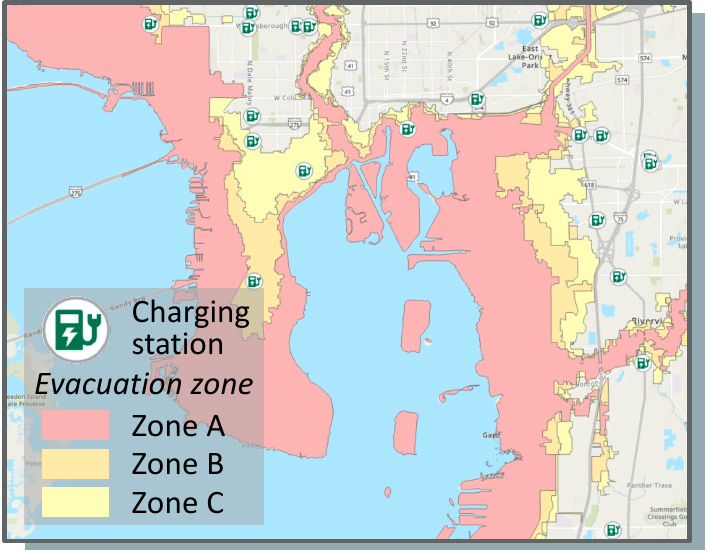}
    \label{fig:b}
}%
\subfloat[]{
\includegraphics[width=0.32\textwidth]{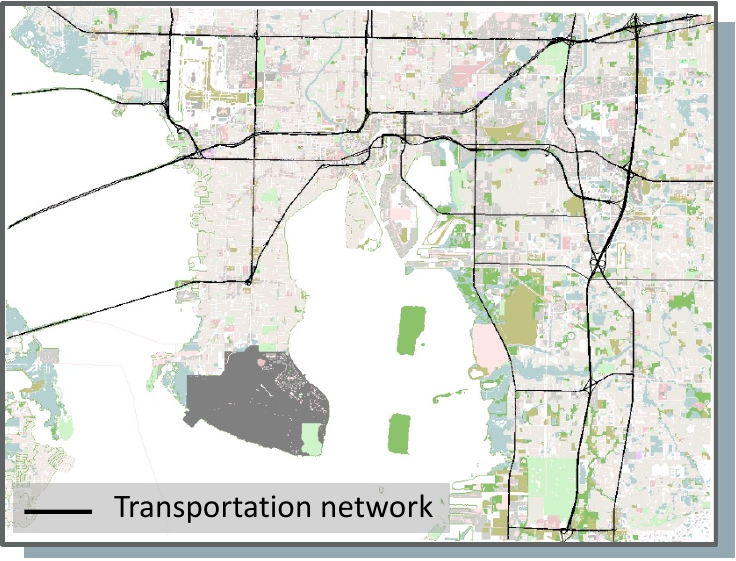}
    \label{fig:c}
}
\caption{Study area in Hillsborough County, Florida: (a) study area; (b) locations of FCSs in study area, marked by green icons; and (c) transportation network used in the SUMO simulator (including 212 nodes and 384 edges).}
\label{fig:case}
\end{figure*}

To evaluate the proposed framework, we use hurricane evacuation as a representative natural-disaster evacuation scenario. A hurricane-prone area in Hillsborough County, Florida, is selected as the case study, as shown in Figure~\ref{fig:case}.
The transportation network and zone classification are obtained from the Hillsborough County Open Data Portal \cite{hillsborough_gis}, while the FCS locations and charger counts are collected from the U.S. Department of Energy \cite{doe_afdc_station}. 
Within this area, there are 20 FCSs, and each charger is assumed to be a fast DC charger with a charging rate of 120 kW. Using the Simulation of Urban Mobility (SUMO) \cite{lopez2018microscopic}, we develop a simulator to reproduce traffic patterns during evacuation. To improve realism, the simulator is calibrated using the real traffic speed data during Hurricane Milton in 2024 by simultaneous perturbation stochastic approximation algorithm \cite{sha2023calibrating}. 
The simulated evacuation spans the 48-hour period prior to hurricane landfall, consistent with empirical evidence that most households evacuate approximately two days before landfall \cite{huang2012household}. 

In this calibrated simulation, the evacuation demand is generated based on household data \cite{census_dhc_2020}. 
According to the average evacuation-order compliance rate reported in \cite{pei2020compound}, we assume the evacuation rate to be 65\%, which means that 65\% of households located in evacuation zones A, B and C evacuate within the simulated horizon.
Their departure times follow a sigmoid curve, meaning that the cumulative departure proportion $F(t)$ by time $t$ is $F(t)=\frac{1}{1+\exp\{-\alpha (t-\beta)\}}$ as adopted in \cite{yazici2008evacuation}.
Here, $\alpha$ controls the steepness of the departure curve and $\beta$ determines the half loading time of the evacuation wave. In the baseline setting, we set $\alpha=0.2$, and $\beta$ is set to 15h, 21h, and 24h for evacuation Zones A, B, and C, respectively.
Based on EV registration data \cite{afdc_ev_registration}, 15\% of all evacuation vehicles are assumed to be EVs. For each EV, the battery capacity is set to 60 kWh, and the initial state of charge at departure is randomly generated from the interval $[0.3,0.8]$. When the battery level falls below 20\%, the EV seeks a reachable FCS in its vicinity for recharging, and charging process continues until the battery is restored to 80\%.

To reflect the evolving hazard condition during evacuation, we model the hazard state $H(t)$ as the remaining time to hurricane landfall. Specifically, hurricane landfall is assumed to occur at 48 h in the simulation horizon, so that $H(t)=48-t$,
where $t$ is the current simulation time in hours.
For station $i$ located in evacuation zone $z(i)$, a zone-specific offset $\delta_{z(i)}$ is introduced to reflect the delayed local impact in different evacuation zones, so that the local hazard state is given by
\[
H_i(t)=H(t)+\delta_{z(i)},
\label{eq:local_hazard_state}
\]
where the offsets are set to 0, 6, and 9 h for Zones A, B, and C, respectively.
The per-capita risk exposure at station $i$ is then specified as an exponential mapping of the local hazard state:
\begin{equation}
R_i(t)=\phi_i(H(t))=
\begin{cases}
\exp\!\left(-H_i(t)/\tau\right), & H_i(t)>0,\\[4pt]
1, & H_i(t)\le 0,
\end{cases}
\label{eq:zone_risk}
\end{equation}
where $\tau$ is the risk growth parameter controlling how fast risk increases as landfall approaches. Under this design, risk exposure increases progressively as the hurricane approaches, and stations in different zones exhibit different risks. To accommodate stochastic charger failures under worsening hazard conditions, the availability of fixed chargers $m_i(t)$ is modeled as a time-varying exogenous process, where the failure probability of each charger increases proportionally with $R_i(t)$.

To supplement the fixed charging infrastructure, six MCTs are introduced into the simulated system. Each MCT carries three mobile chargers, with a total onboard battery capacity of 3000 kWh per truck. In the allocation module, MCT deployment decisions are updated at decision epochs $\mathcal{T}^{\mathrm{dec}}=\{\tau_1,\tau_2,\dots,\tau_M\}$ rather than at every simulation step. Once an MCT arrives at its assigned FCS, it remains on site for a 2-hour service period before becoming available for reallocation. Accordingly, the decision-epoch sequence is defined on a coarser timescale than the simulation clock and reflects the practical service cycle of mobile charging support. Under decentralized execution, each MCT normally observes only a local subset of FCSs. The periodic observation augmentation is applied every three decision epochs, i.e., $\Delta_{\mathrm{aug}}=3$, so that each MCT temporarily receives an expanded network view.

\subsection{Experiment Setup}
\subsubsection{Configuration of Allocation Module}
For the allocation module in Problem I,  
both actor and critic network use the relational station encoder described in Section~\ref{sec:encoder_PPO}, with two attention heads and a model dimension of $d_{\mathrm{model}}=64$. The actor outputs a categorical policy distribution over the candidate FCSs through a softmax layer, while the critic outputs a scalar value estimate through a linear output layer.
During offline policy pre-training, the discount factor is set to $\gamma=0.99$, the learning rate is set to $0.005$, and the PPO clipping parameter is set to $\epsilon=0.15$. During online deployment, the retrieval module selects the top $K_{\mathrm{ret}}=256$ most similar records from the experience bank for actor adaptation.

\subsubsection{Configuration of Routing Module}
For the routing module in Problem II, the STPM adopts an encoder-decoder architecture with eight attention heads and a model dimension of $d_{\mathrm{model}}=64$. The model is trained using simulator-generated edge-level travel time sequences under 1200 evacuation scenarios, each produced with a different random seed so as to reflect the variability of evacuation traffic patterns induced by stochastic evacuee movements. Traffic states are aggregated at 5-minute intervals. Both the input history window and the prediction window are set to one hour, corresponding to 12 time steps. During training, the learning rate is set to $0.001$. It should be noted that although STPM is trained on simulation data in this study, the proposed prediction framework can also be trained using real-world traffic observations when available.

\subsubsection{Model Comparisons}
The proposed ARMD is evaluated against the following benchmark and ablation models:
\begin{itemize}
    \item \textbf{No-MCT deployment (No-MCT):} No mobile charging trucks are deployed, and the evacuation charging system relies entirely on the fixed charging infrastructure.
    \item \textbf{Greedy:} At each decision epoch, each MCT is assigned to the observable FCS with the highest station-level risk exposure, i.e., the largest value of $R_i(t)Q_i(t)$.
     \item \textbf{Centralized allocation (CA):} CA adopts a centralized policy to output the joint actions of all MCTs simultaneously. 
     This policy uses the same model dimension and training settings as the MAPPO-based allocation module in ARMD, but adopts centralized action generation.
    \item \textbf{Offline fixed optimization (OF-MIP):} 
    MCT deployment is formulated as an MIP model, with the detailed formulation provided in Appendix~\ref{MILP_model}. OF-MIP solves this MIP once over the entire evacuation horizon using forecasted charging demand and traffic dynamics. It represents an open-loop benchmark, where the MCT deployment schedule is optimized once before evacuation and then applied without online re-optimization. 
    \item \textbf{Rolling-horizon MIP (RH-MIP):} 
    RH-MIP uses the same MIP formulation as OF-MIP, but implements it in a closed-loop rolling-horizon manner. At each decision epoch, the model is re-solved using the current FCS and MCT states, and forecasted profiles over a finite planning horizon. Only the decisions for the current epoch are implemented, and the optimization is re-solved at the next epoch after the system state is updated.
    \item \textbf{ARMD-NF: } This ablation removes online actor fine-tuning. The allocation policy is executed directly using the offline pre-trained model. 
    \item \textbf{ARMD-NR: } This ablation removes online route updating. The routing decision is determined without rolling-horizon re-optimization during execution. 
    \item \textbf{Ours (ARMD): } This adopts the full proposed two-stage online allocation and routing framework.
\end{itemize}

\subsubsection{Experimental Scenarios}
To evaluate the performance and resilience of the proposed ARMD framework, we design three types of evacuation scenarios that represent different operational conditions and unexpected disruptions during evacuation. The scenario names defined below are used in the subsequent tables and figures. These scenarios are used to assess whether the proposed method can adapt effectively to varying demand patterns and infrastructure conditions.

\begin{itemize}
    \item \textbf{Base scenario:} 
    The \textit{training environment} follows the same setting as in the model training, but different seeds are used to generate different system evolutions.

    \item \textbf{Evacuation demand perturbation scenarios:} 
    In these scenarios, the evacuation demand deviates from that in the model training setting. We consider three cases: \textit{higher evacuation participation}, where the evacuation rate increases to 75\% of households; \textit{flatter departure}, where the departure curve of evacuees is flatter with $\alpha=0.15$; and \textit{concentrated departure}, where the departure curve is more concentrated with $\alpha=0.25$.

    \item \textbf{Infrastructure failure scenarios:}
    These scenarios represent infrastructure degradation under extreme weather conditions. Two types of failures are considered. The first type is fixed charging infrastructure failure, where a fraction of FCSs becomes unavailable, representing reduced charging supply. We consider station failure probabilities of \textit{15\%}, \textit{30\%}, and \textit{45\%}.
    The second type is road link failure, including \textit{reduced capacity}, where a subset of road links has reduced capacity; \textit{two components}, where complete disruption of these links splits the transportation network into two disconnected components; and \textit{three components}, where additional link disruptions further divide the network into three disconnected components. These scenarios are intended to capture road network degradation caused by road obstructions or closures.
\end{itemize}

To isolate the contribution of the online routing module, we compare the travel time of MCT reallocation trips across the test scenarios under ARMD and ARMD-NR. These trips are classified into long and short moves using a 60-min threshold.

\subsubsection{Evaluation Metrics}
We define three performance metrics to evaluate the proposed method and the baseline models.
\begin{itemize}
    \item \textbf{Average risk exposure (ARE):} ARE measures the average queue-induced risk exposure for each EV evacuee over the entire horizon 
    \begin{equation}
        ARE = \frac{1}{N^{\mathrm{evac}}} \sum_{t \in \mathcal{T}} \sum_{i \in \mathcal{F}} R_i(t)\,Q_i(t),
    \end{equation}
    where $N^{\mathrm{evac}}$ denotes the total number of EV evacuees on the simulation horizon.
    
    \item \textbf{Peak station-level risk exposure (PSRE):} PSRE captures the worst station-time risk condition: 
    \begin{equation}
        PSRE = \max_{t \in \mathcal{T},\ i \in \mathcal{F}} R_i(t)\,Q_i(t).
    \end{equation}
    
    \item \textbf{Average station-level risk exposure in the late stage (ASRE-L):} Since the hazard intensifies as hurricane landfall approaches, risk exposure in the late stage of evacuation is of particular concern. We therefore define
    \begin{equation}
        ASRE\text{-}L =
        \frac{
        \sum_{t \in \mathcal{T}^{\mathrm{late}}}\sum_{i \in \mathcal{F}} R_i(t)\,Q_i(t)
        }{
        \sum_{t \in \mathcal{T}^{\mathrm{late}}}\sum_{i \in \mathcal{F}} \mathbf{1}\!\left(R_i(t)\,Q_i(t)>0\right)
        },
    \end{equation}
    where $\mathcal{T}^{\mathrm{late}} \subseteq \mathcal{T}$ denotes the set of time steps in the late stage, taken here as the last 12 hours of the evacuation horizon, and $\mathbf{1}(\cdot)$ is the indicator function. 
\end{itemize}

All experiments were conducted on the Ubuntu 22.04.4 LTS server with an AMD Ryzen 9 7950X, 128 GB RAM, and NVIDIA RTX 6000 Ada, utilizing the Pytorch library.

\subsection{Result Analysis}
\subsubsection{Queue Dynamics}
\begin{figure}[htbp]
    \centering
    \subfloat[Without MCT deployment]{
        \includegraphics[width=0.48\linewidth]{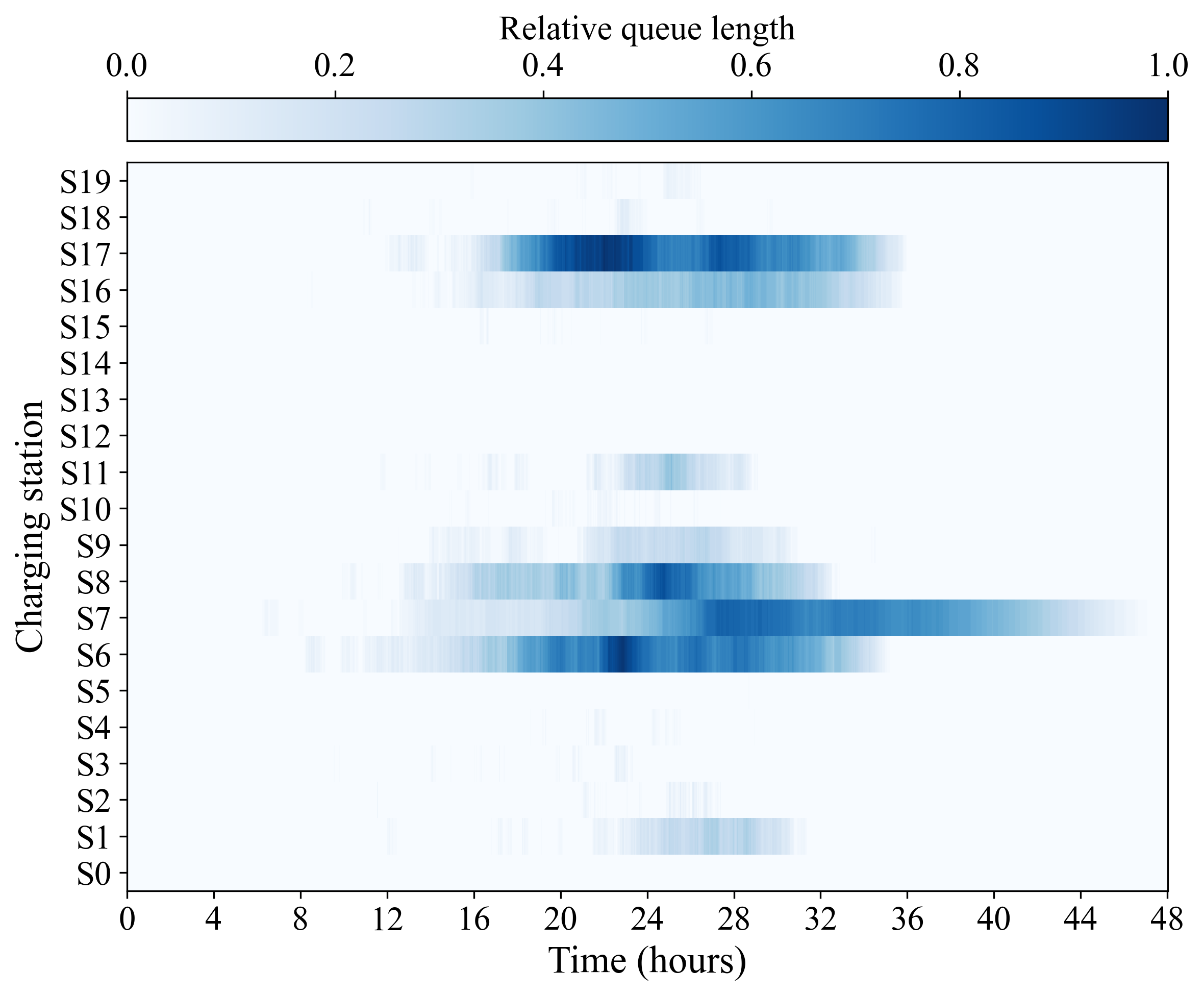}
        \label{fig:queue_heatmap_nomct}
    }
    \subfloat[With dynamic MCT allocation using ARMD]{
        \includegraphics[width=0.48\linewidth]{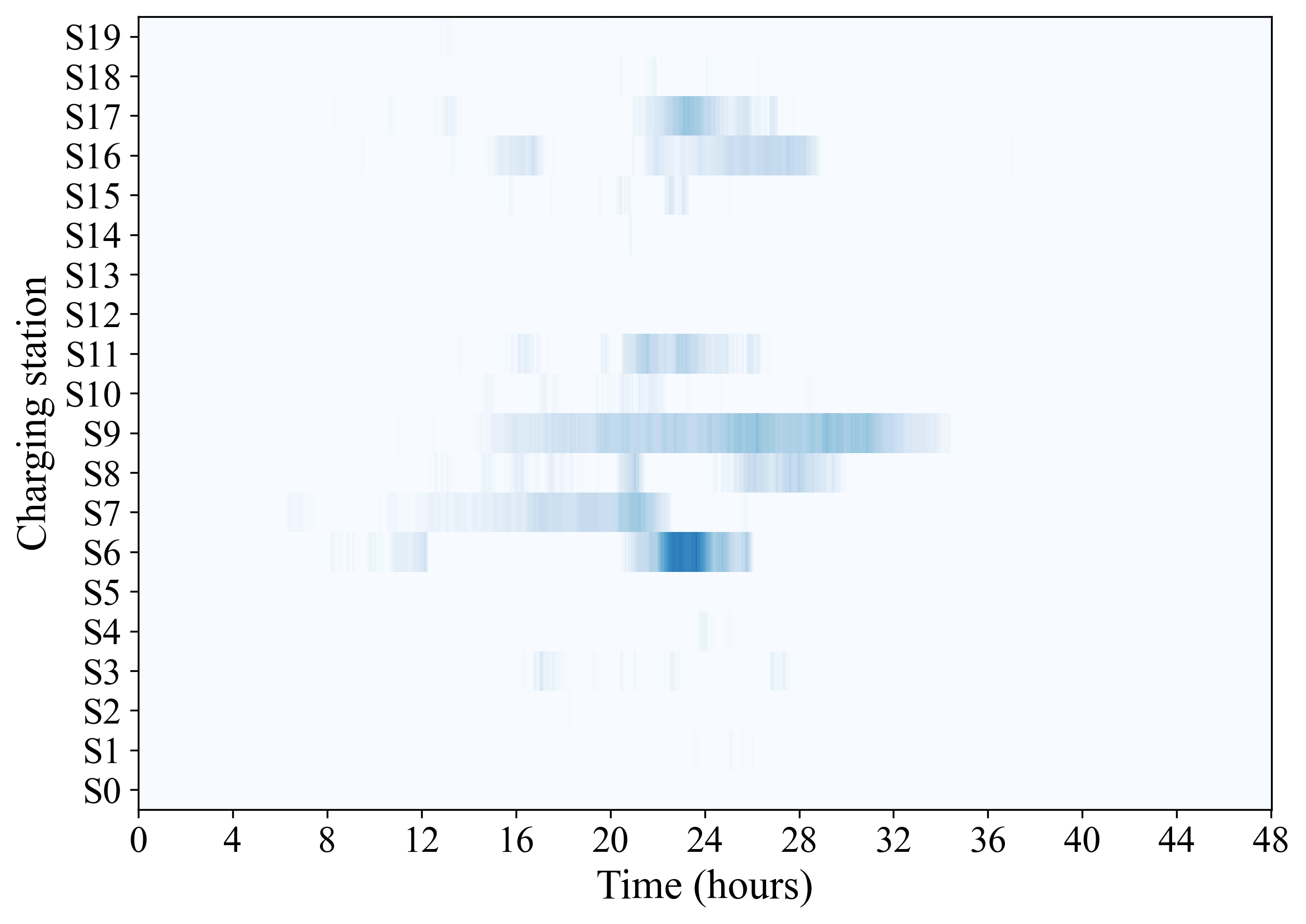}
        \label{fig:queue_heatmap_s2oa}
    }
    \caption{Station-level queue heatmaps over time under the base scenario.}
    \label{fig:queue_heatmap}
\end{figure}

We first examine the spatiotemporal dynamics of charging queues during evacuation. Figure~\ref{fig:queue_heatmap} compares the queue heatmaps over time under the base scenario without MCT deployment and with dynamic MCT allocation by ARMD. It can be observed that queue buildup is highly uneven across FCSs and evolves substantially over time under the no-MCT case, with severe congestion concentrated at a subset of stations during specific periods. In Figure~\ref{fig:queue_heatmap}(b), the spatiotemporal pattern remains dynamic, 
but the queue intensity at major hotspot stations is reduced and the congestion becomes less concentrated and less persistent over time.
This indicates that evacuation charging demand is spatially heterogeneous and temporally dynamic, and adaptive MCT allocation can help mitigate localized queue accumulation.

\begin{table*}[htbp]
\centering
\footnotesize
\setlength{\tabcolsep}{1pt}
\caption{Performance comparison under the base scenario and evacuation demand perturbations. The best results are shown in bold and highlighted, and the second-best results are underlined.}
\label{tab:performance_demand}
\begin{tabular}{>{\raggedright\arraybackslash}p{1.6cm} l c c c c c c c c}
\toprule
\multicolumn{10}{c}{\textit{Base scenario}} \\
\cmidrule(){3-10}
 &  & No-MCT & Greedy & CA & OF-MIP & RH-MIP & ARMD-NF & ARMD-NR & ARMD \\
\midrule
\multirow{3}{1.6cm}{Training environment}
& ARE      & \num{2.6\pm0.6} & \num{2.1\pm0.5} & \best{1.2\pm0.6} & \num{1.4\pm0.2} & \second{1.4\pm0.1} & \num{1.7\pm0.4} & \num{1.5\pm0.6} & \num{1.4\pm0.6} \\
& PSRE     & \num{21.8\pm4.6} & \num{21.4\pm1.0} & \best{17.5\pm4.7} & \num{18.5\pm1.9} & \second{17.9\pm0.5} & \num{20.2\pm1.0} & \num{20.5\pm1.3} & \num{19.5\pm2.3} \\
& ASRE-L   & \num{13.3\pm5.5} & \num{14.1\pm3.6} & \second{5.0\pm6.2} & \num{8.4\pm1.7} & \num{7.6\pm0.8} & \num{6.6\pm5.9} & \num{8.0\pm4.2} & \best{4.7\pm5.8} \\
\midrule

\multicolumn{10}{c}{\textit{Evacuation demand perturbations}} \\
\cmidrule(){3-10}
 &  & No-MCT & Greedy & CA & OF-MIP & RH-MIP & ARMD-NF & ARMD-NR & ARMD \\
\midrule
\multirow{3}{1.6cm}{Flatter departure}
& ARE      & \num{2.5\pm0.4} & \num{1.9\pm0.4} & \num{1.6\pm1.0} & \num{1.0\pm0.1} & \num{0.8\pm0.1} & \num{0.8\pm0.2} & \second{0.7\pm0.3} & \best{0.7\pm0.1} \\
& PSRE     & \num{23.3\pm2.3} & \num{19.9\pm2.3} & \num{17.9\pm7.4} & \num{19.9\pm2.1} & \num{15.5\pm2.0} & \num{15.5\pm1.9} & \best{14.7\pm2.5} & \second{15.3\pm2.6} \\
& ASRE-L   & \num{14.9\pm5.4} & \num{12.6\pm3.0} & \num{10.9\pm9.6} & \num{9.7\pm1.5} & \num{5.0\pm0.7} & \num{2.7\pm2.0} & \best{2.0\pm2.4} & \second{2.6\pm3.1} \\
\midrule

\multirow{3}{1.6cm}{Concentrated departure}
& ARE      & \num{3.3\pm0.5} & \num{2.6\pm1.0} & \num{2.5\pm0.9} & \num{2.1\pm0.2} & \num{2.0\pm0.1} & \num{2.0\pm0.7} & \second{1.8\pm0.7} & \best{1.7\pm0.7} \\
& PSRE     & \num{23.4\pm2.4} & \num{23.2\pm4.0} & \num{25.0\pm4.9} & \num{22.3\pm0.1} & \num{23.0\pm1.2} & \num{22.0\pm2.7} & \second{21.2\pm1.9} & \best{19.8\pm3.8} \\
& ASRE-L   & \num{12.6\pm6.0} & \num{12.9\pm6.9} & \num{8.5\pm9.2} & \num{9.9\pm0.9} & \num{9.9\pm2.7} & \num{9.0\pm7.6} & \second{3.9\pm5.5} & \best{3.5\pm5.0} \\
\midrule

\multirow{3}{1.6cm}{Higher evacuation participation}
& ARE      & \num{5.8\pm0.2} & \num{3.6\pm0.6} & \num{4.7\pm0.3} & \num{4.1\pm0.5} & \num{2.8\pm0.1} & \second{2.6\pm0.1} & \num{3.2\pm0.1} & \best{2.2\pm0.3} \\
& PSRE     & \num{45.7\pm19.9} & \num{37.2\pm10.3} & \num{42.2\pm5.5} & \num{35.4\pm5.8} & \num{38.9\pm2.1} & \num{37.0\pm11.1} & \best{30.7\pm9.1} & \second{31.7\pm5.2} \\
& ASRE-L   & \num{11.4\pm3.7} & \num{14.1\pm5.6} & \num{12.4\pm4.3} & \num{9.8\pm1.4} & \best{4.9\pm0.6} & \num{8.3\pm6.2} & \num{6.5\pm5.6} & \second{6.2\pm4.6} \\
\bottomrule
\end{tabular}
\end{table*}

\subsubsection{Results under Evacuation Demand Perturbations}
Table~\ref{tab:performance_demand} summarizes the results under base scenario and evacuation demand perturbation scenarios. 
Compared with the No-MCT baseline, all MCT-enabled methods exhibit lower risks, confirming that mobile charging support helps mitigate queue-induced risk exposure during evacuation.
Among the perturbed scenarios, higher evacuation participation is the most challenging, yielding the highest ARE, PSRE, and ASRE-L. 

Under the base scenario, CA achieves the lowest ARE and PSRE, while ARMD obtains the lowest ASRE-L. This is generally consistent with expectation, since CA makes decisions based on global information and can therefore achieve coordinated system-level allocation under the training environment. When the demand pattern deviates from the training environment, ARMD maintains strong performance across all demand-perturbation scenarios, indicating stable generalization under evacuation demand perturbations.
The advantage of ARMD is particularly evident under the flatter-departure scenario, where it reduces ARE by $71.1\%$ relative to No-MCT. By comparison, Greedy, CA, and RH-MIP reduce ARE by $21.5\%$, $35.0\%$, and $69.1\%$, respectively. Similar benefits are also observed under the concentrated-departure and higher-participation scenarios, where ARMD reduces ARE by $49.7\%$ and $61.4\%$, respectively. We attribute this advantage to the decentralized and adaptive design of ARMD, which enables each MCT to respond more flexibly to local demand variations.

We also find that RL-based methods consistently outperform the Greedy baseline, with the most pronounced gap observed in ASRE-L. The ASRE-L under the Greedy method is often close to or worse than the No-MCT baseline. 
This indicates that reacting only to immediate urgency may relieve short-term pressure but fails to mitigate residual risk accumulation near the end of evacuation, when hazard exposure becomes most critical. In contrast, RL-based methods better account for the longer-term system effect of current allocation decisions, thereby achieving improved late-stage performance.

\begin{table*}[htbp]
\centering
\footnotesize
\setlength{\tabcolsep}{1pt}
\caption{Performance comparison under infrastructure failure scenarios. The best results are shown in bold and highlighted, and the second-best results are underlined.}
\label{tab:performance_failure}
\begin{tabular}{>
{\raggedright\arraybackslash}p{1.6cm} l c c c c c c c c}
\toprule
\multicolumn{10}{c}{\textit{Fixed charging infrastructure failure scenarios}} \\
\cmidrule(){3-10}
 &  & No-MCT & Greedy & CA & OF-MIP & RH-MIP & ARMD-NF & ARMD-NR & ARMD \\
\midrule

\multirow{3}{1.6cm}{Station failure probability = 15\%}
& ARE    & \numnum{3.9\pm0.2} & \numnum{3.0\pm0.7} & \numnum{2.6\pm0.7} & \numnum{2.9\pm0.3} & \numnum{2.8\pm0.2} & \numnum{2.8\pm0.8} & \secondsecond{2.6\pm1.2} & \bestbest{2.2\pm1.1} \\
& PSRE   & \numnum{26.5\pm2.7} & \numnum{23.7\pm1.7} & \numnum{23.3\pm2.6} & \numnum{26.8\pm1.8} & \numnum{25.5\pm1.3} & \secondsecond{21.7\pm3.9} & \numnum{22.6\pm5.6} & \bestbest{20.6\pm3.8} \\
& ASRE-L & \numnum{16.1\pm2.0} & \numnum{15.7\pm2.3} & \secondsecond{9.6\pm5.6} & \numnum{14.6\pm1.4} & \numnum{12.6\pm2.4} & \numnum{11.0\pm5.8} & \numnum{11.0\pm10.1} & \bestbest{9.5\pm7.8} \\
\midrule

\multirow{3}{1.6cm}{Station failure probability = 30\%}
& ARE    & \numnum{5.7\pm0.2} & \numnum{4.4\pm0.3} & \numnum{4.2\pm0.8} & \numnum{5.0\pm0.5} & \numnum{4.1\pm0.2} & \numnum{4.4\pm0.9} & \secondsecond{3.5\pm1.1} & \bestbest{3.2\pm1.0} \\
& PSRE   & \numnum{31.7\pm2.2} & \numnum{32.0\pm2.6} & \numnum{31.3\pm1.7} & \numnum{38.0\pm5.3} & \numnum{52.3\pm3.2} & \numnum{30.0\pm2.2} & \secondsecond{26.3\pm4.9} & \bestbest{24.9\pm4.8} \\
& ASRE-L & \numnum{19.8\pm1.9} & \numnum{22.6\pm2.7} & \numnum{16.8\pm6.8} & \numnum{24.6\pm4.9} & \numnum{27.3\pm1.0} & \numnum{19.4\pm8.7} & \secondsecond{11.6\pm11.8} & \bestbest{6.6\pm10.4} \\
\midrule

\multirow{3}{1.6cm}{Station failure probability = 45\%}
& ARE    & \numnum{10.5\pm0.6} & \numnum{6.4\pm0.9} & \numnum{7.7\pm1.9} & \numnum{7.6\pm0.6} & \numnum{6.7\pm0.6} & \numnum{4.8\pm2.2} & \secondsecond{4.4\pm1.5} & \bestbest{4.0\pm1.9} \\
& PSRE   & \numnum{49.0\pm3.4} & \numnum{49.5\pm6.4} & \numnum{48.4\pm9.3} & \numnum{45.3\pm7.5} & \numnum{91.1\pm8.3} & \numnum{36.8\pm12.9} & \secondsecond{35.2\pm10.0} & \bestbest{33.4\pm13.5} \\
& ASRE-L & \numnum{37.3\pm17.8} & \numnum{43.6\pm9.1} & \numnum{38.6\pm2.4} & \numnum{37.1\pm7.2} & \numnum{33.9\pm4.8} & \numnum{23.5\pm20.2} & \bestbest{19.5\pm17.3} & \secondsecond{21.1\pm17.8} \\
\midrule

\multicolumn{10}{c}{\textit{Road link failure scenarios}} \\
\cmidrule(){3-10}
 &  & No-MCT & Greedy & CA & OF-MIP & RH-MIP & ARMD-NF & ARMD-NR & ARMD \\
\midrule

\multirow{3}{1.6cm}{Road failure: reduced capacity}
& ARE    & \numnum{2.8\pm0.4} & \numnum{2.6\pm0.3} & \numnum{2.1\pm0.5} & \numnum{2.5\pm0.6} & \numnum{1.8\pm0.2} & \secondsecond{1.8\pm0.7} & \numnum{2.0\pm0.7} & \bestbest{1.7\pm0.8} \\
& PSRE   & \numnum{25.2\pm3.2} & \numnum{27.2\pm3.1} & \secondsecond{21.8\pm3.4} & \numnum{26.6\pm2.4} & \numnum{23.9\pm1.2} & \bestbest{21.7\pm3.7} & \numnum{24.4\pm4.0} & \numnum{22.4\pm2.8} \\
& ASRE-L & \numnum{14.1\pm2.5} & \numnum{17.9\pm4.2} & \numnum{9.4\pm6.1} & \numnum{14.5\pm4.0} & \numnum{11.8\pm2.2} & \secondsecond{8.6\pm5.9} & \numnum{9.4\pm8.7} & \bestbest{8.5\pm7.1} \\
\midrule

\multirow{3}{1.6cm}{Road failure: two components}
& ARE    & \numnum{3.3\pm0.4} & \numnum{2.6\pm0.5} & \numnum{2.4\pm0.2} & \numnum{2.6\pm0.2} & \secondsecond{2.3\pm0.3} & \numnum{2.4\pm0.3} & \numnum{2.5\pm0.9} & \bestbest{1.5\pm0.3} \\
& PSRE   & \numnum{27.0\pm5.0} & \numnum{29.6\pm3.2} & \numnum{30.7\pm5.1} & \numnum{39.4\pm7.1} & \numnum{24.9\pm2.4} & \numnum{28.2\pm2.6} & \bestbest{25.9\pm5.5} & \secondsecond{26.5\pm4.3} \\
& ASRE-L & \numnum{8.3\pm1.1} & \numnum{8.6\pm1.0} & \numnum{6.4\pm0.7} & \numnum{8.0\pm0.7} & \secondsecond{5.8\pm2.5} & \numnum{6.3\pm3.1} & \numnum{7.8\pm2.3} & \bestbest{3.6\pm2.3} \\
\midrule

\multirow{3}{1.6cm}{Road failure: three components}
& ARE    & \numnum{2.6\pm0.9} & \numnum{2.2\pm0.9} & \numnum{1.6\pm0.8} & \numnum{2.3\pm1.4} & \numnum{1.6\pm0.7} & \numnum{1.2\pm0.8} & \secondsecond{1.1\pm0.2} & \bestbest{1.0\pm0.2} \\
& PSRE   & \numnum{48.1\pm20.1} & \numnum{42.0\pm16.3} & \numnum{31.1\pm12.0} & \numnum{52.4\pm32.7} & \numnum{39.6\pm19.6} & \secondsecond{25.0\pm9.4} & \numnum{27.1\pm7.0} & \bestbest{24.5\pm4.7} \\
& ASRE-L & \numnum{9.1\pm3.7} & \numnum{8.0\pm3.3} & \numnum{7.2\pm7.1} & \numnum{7.9\pm6.6} & \numnum{5.4\pm2.9} & \numnum{4.3\pm2.5} & \secondsecond{3.5\pm1.0} & \bestbest{3.0\pm0.7} \\
\bottomrule
\end{tabular}
\end{table*}

\subsubsection{Results for Infrastructure Failure Scenarios}
Table~\ref{tab:performance_failure} shows the performance of all methods under infrastructure failure scenarios. For all these scenarios, ARMD reduces ARE by $39.3\%$--$60.5\%$ relative to No-MCT, indicating its overall effectiveness in mitigating queue-induced risk under degraded infrastructure conditions.

\begin{figure*}[!t]
\centering
\subfloat[Flatter departure]{
    \includegraphics[width=0.33\textwidth]{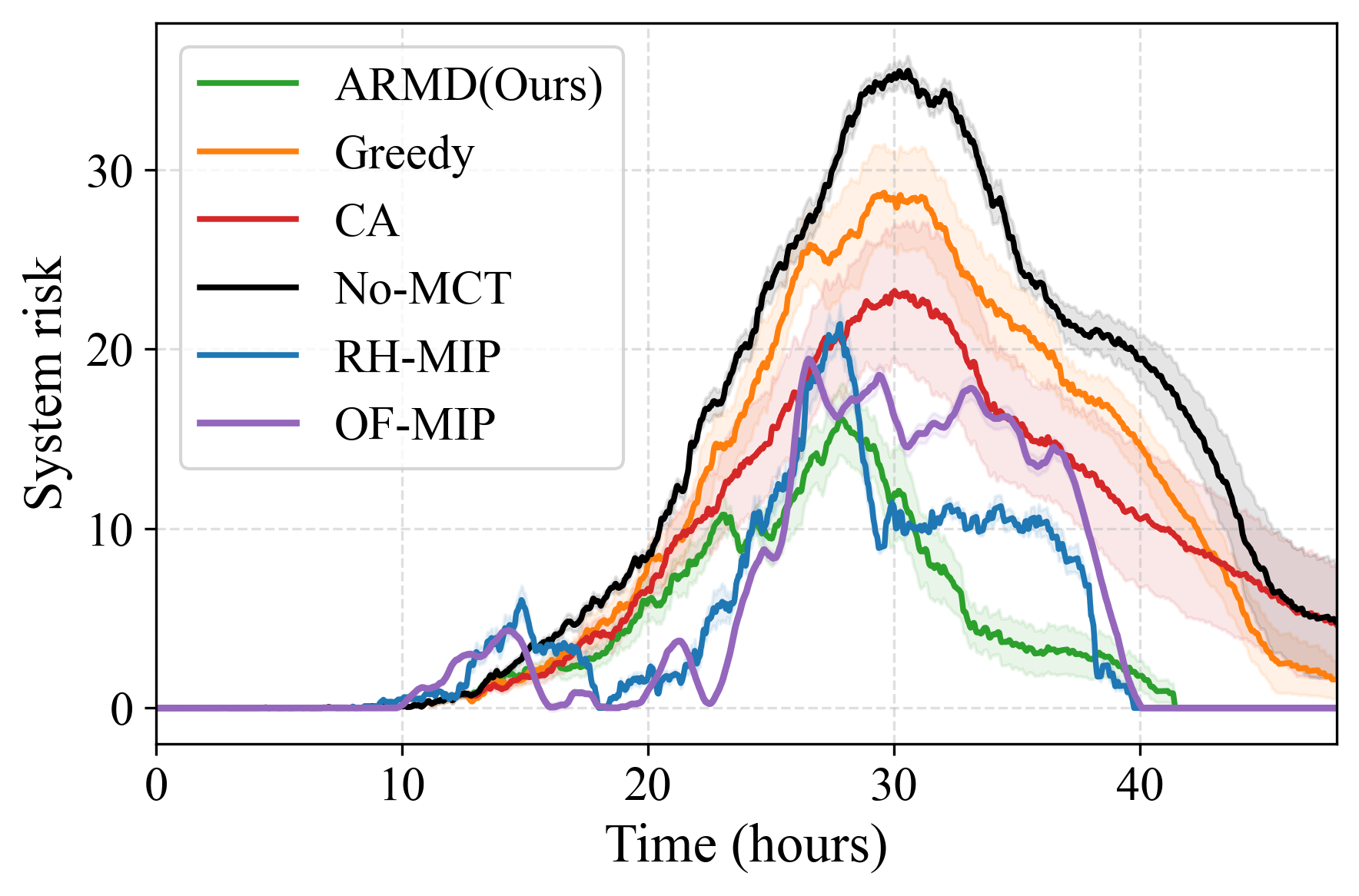}
    \label{fig:gradual}
}
\hspace{-3mm}
\subfloat[Concentrated departure]{
    \includegraphics[width=0.33\textwidth]{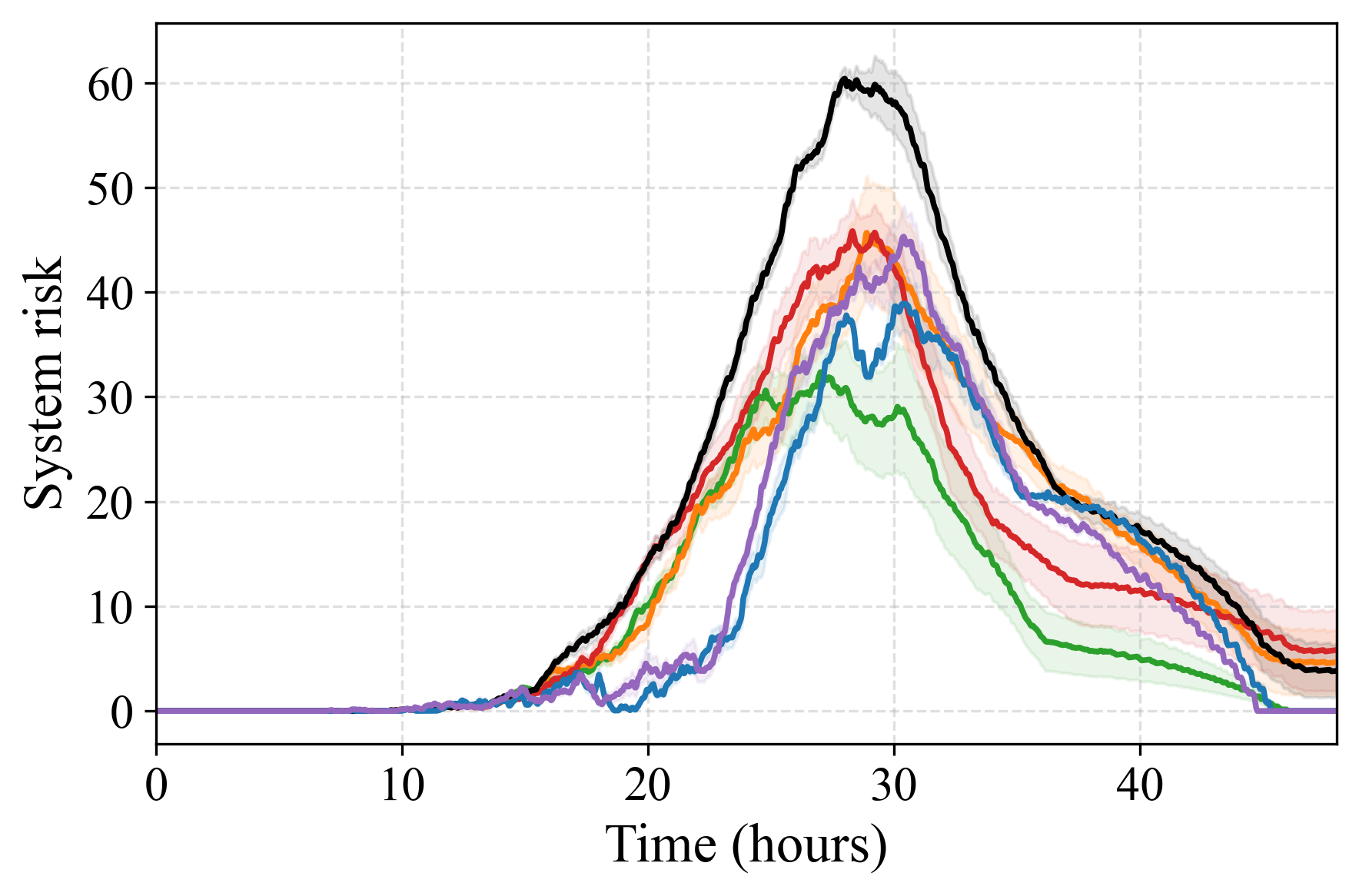}
    \label{fig:con}
}
\hspace{-3mm}
\subfloat[Higher evacuation participation]{
    \includegraphics[width=0.33\textwidth]{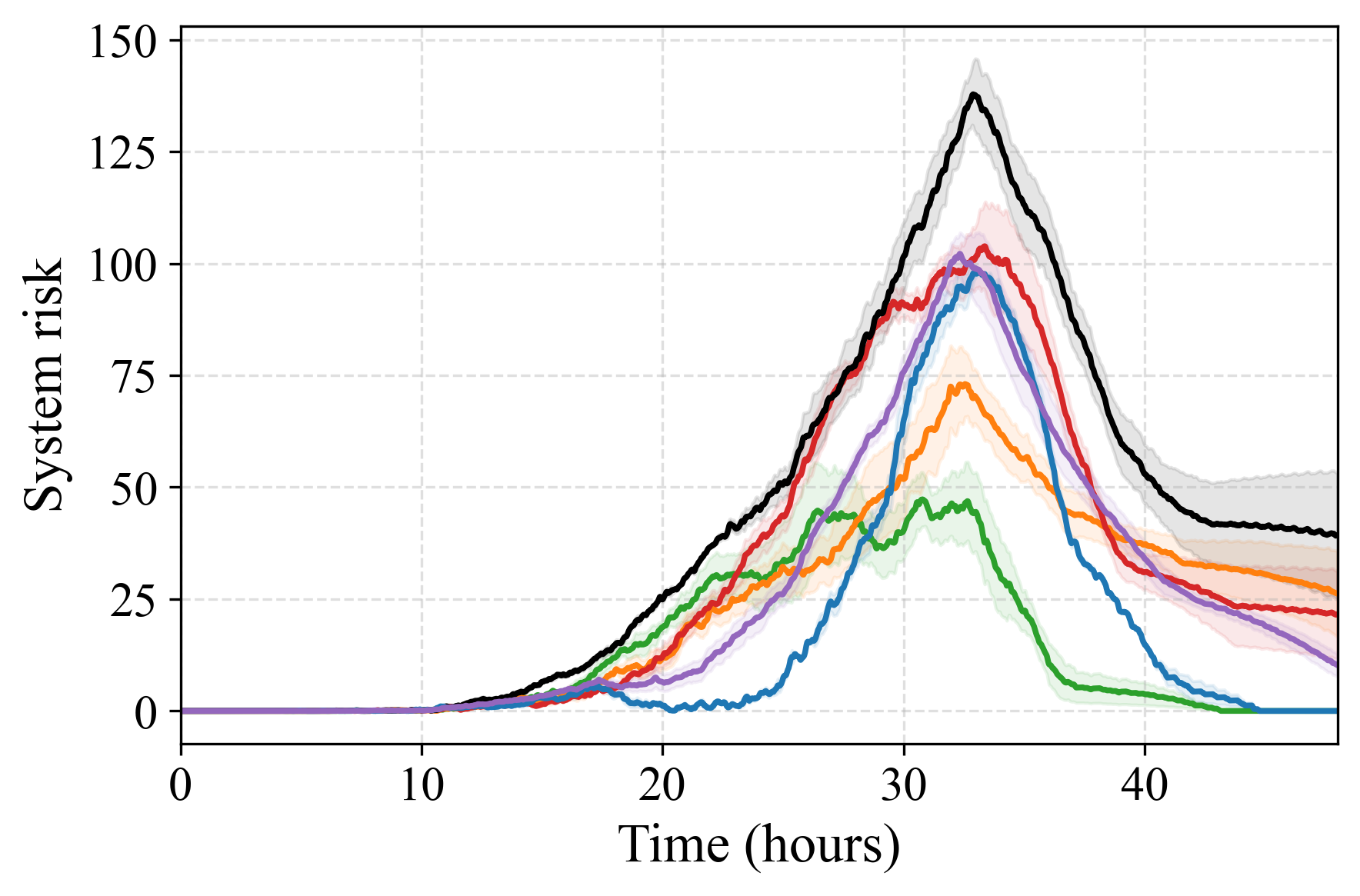}
    \label{fig:surge}
}
\\[1ex]

\subfloat[Station failure
probability = $15\%$]{
    \includegraphics[width=0.33\textwidth]{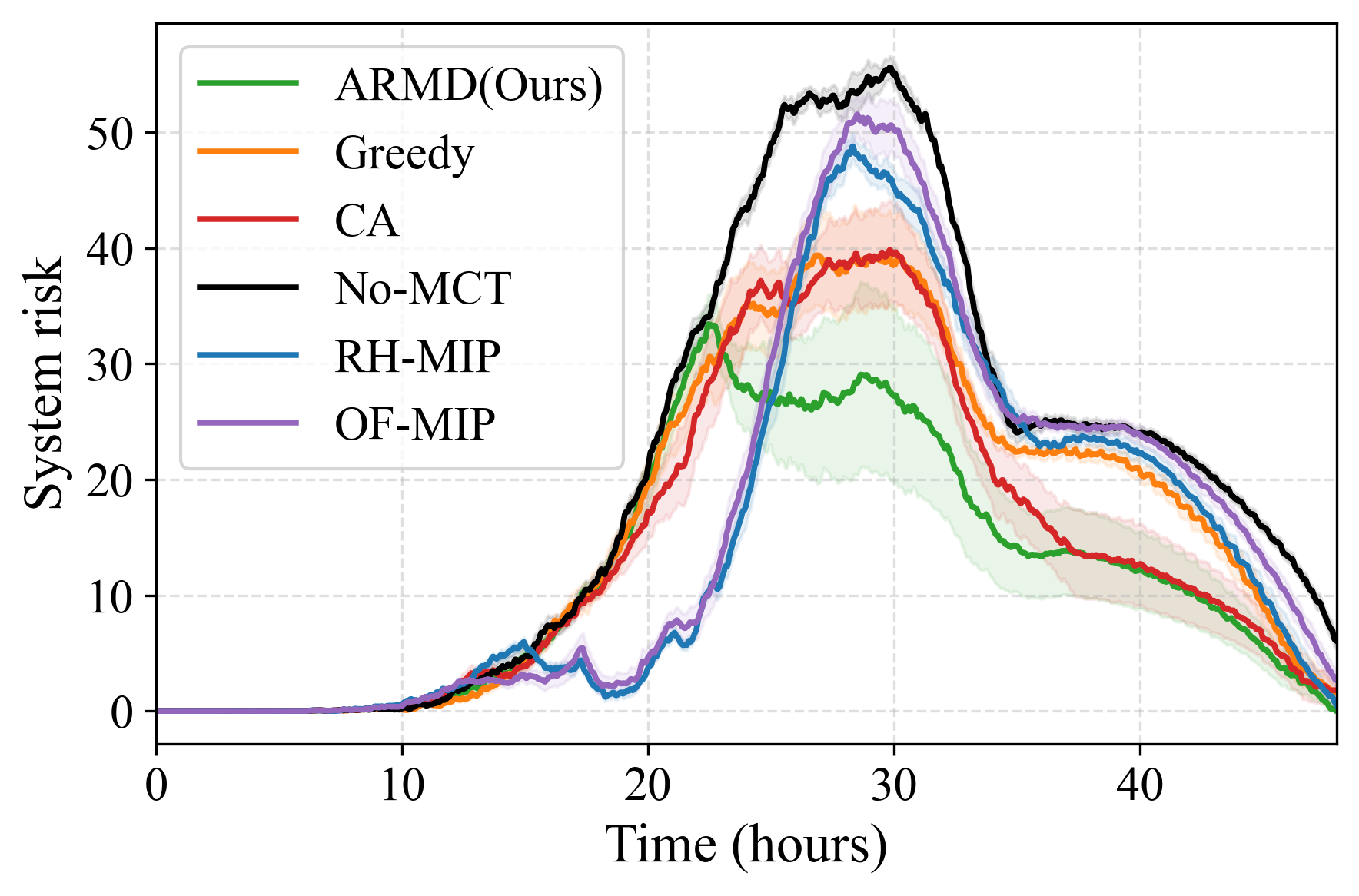}
    \label{fig:failure_risk_15}
}
\hspace{-3mm}
\subfloat[Station failure
probability = $30\%$]{
    \includegraphics[width=0.33\textwidth]{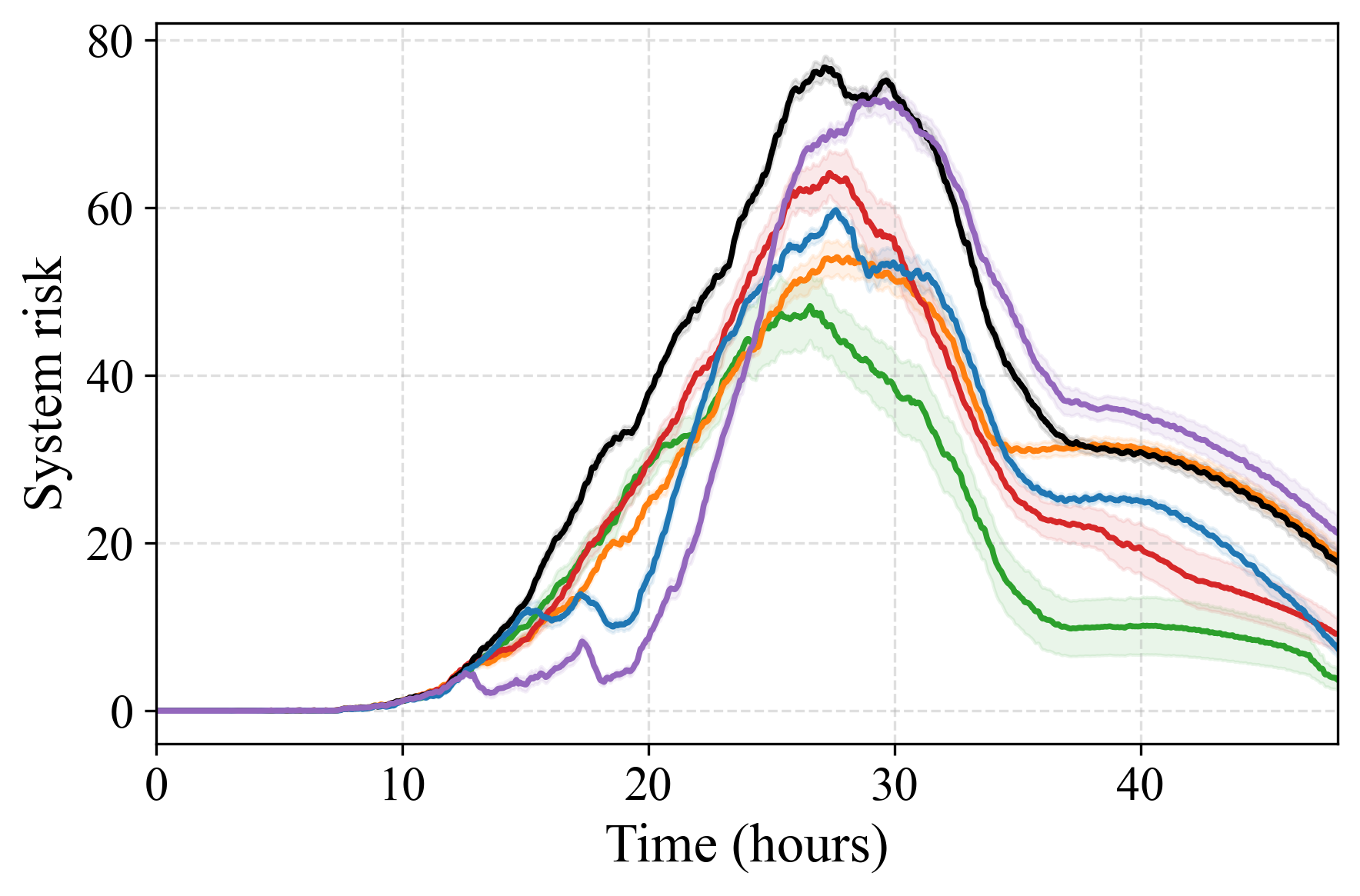}
    \label{fig:failure_risk_30}
}
\hspace{-3mm}
\subfloat[Station failure
probability = $45\%$]{
    \includegraphics[width=0.33\textwidth]{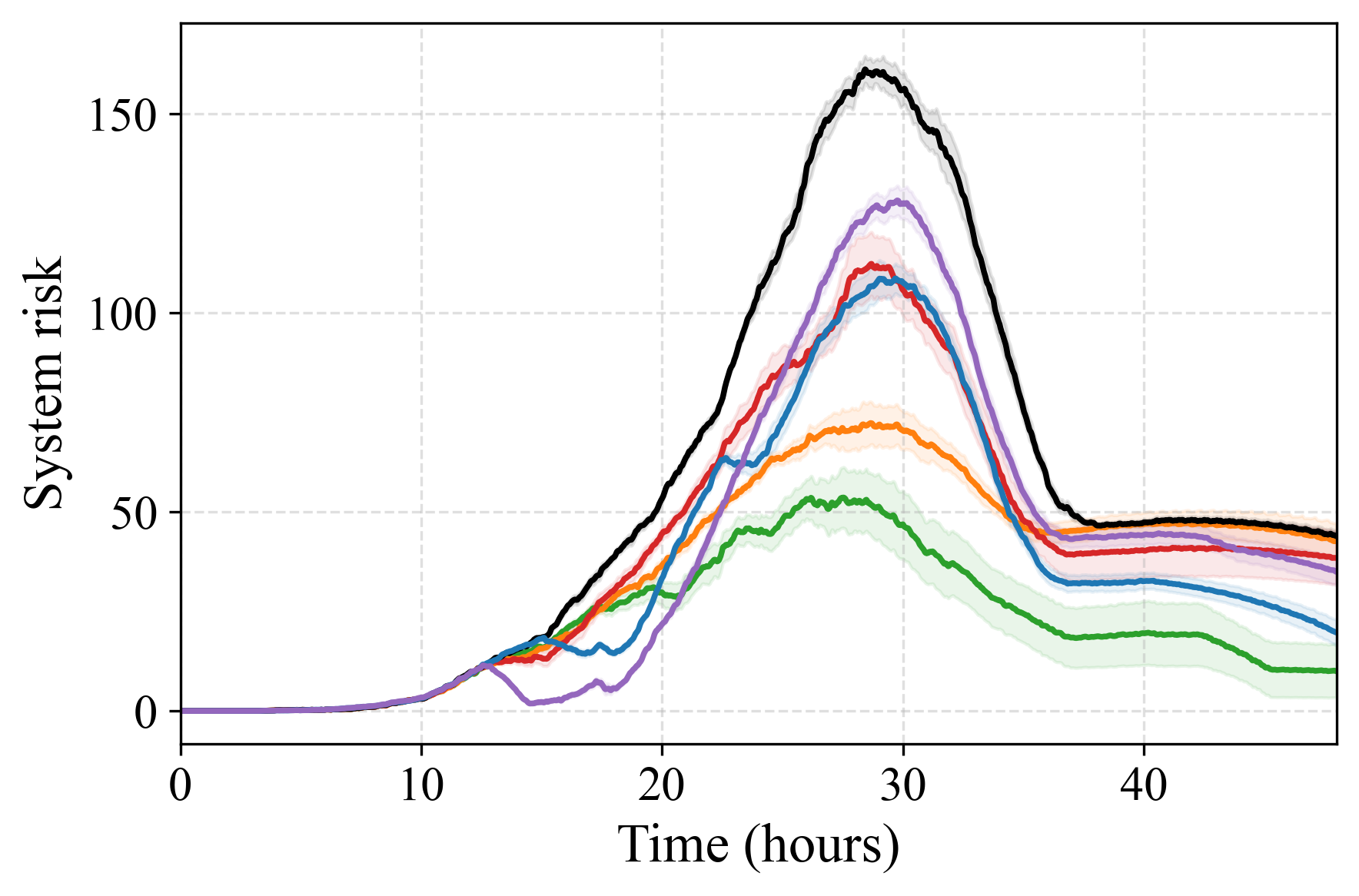}
    \label{fig:failure_risk_45}
}
\\[1ex] 

\subfloat[Road failure - reduced capacity]{
    \includegraphics[width=0.33\textwidth]{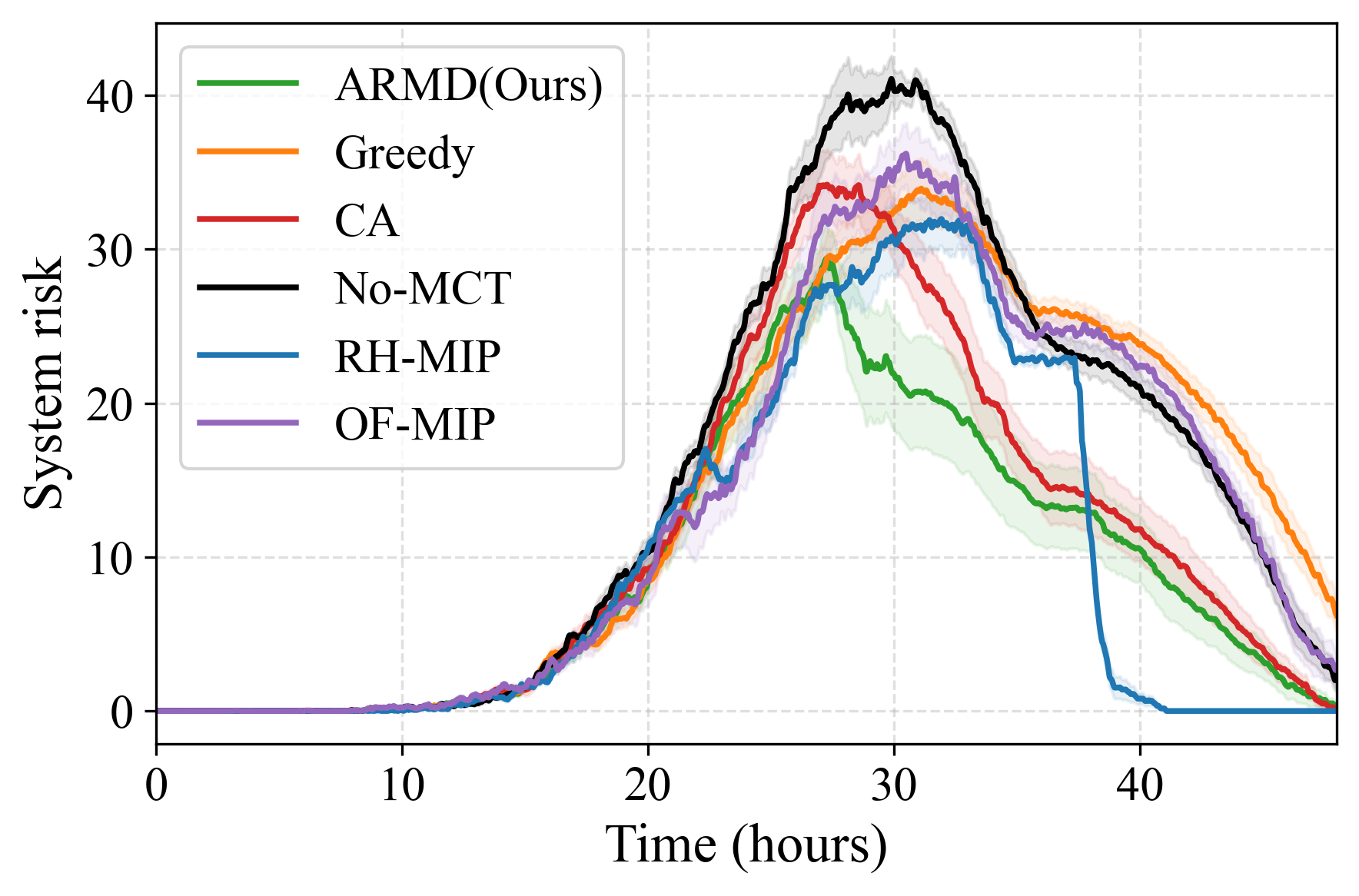}
    \label{fig:road1}
}
\hspace{-3mm}
\subfloat[Road failure - two components]{
    \includegraphics[width=0.33\textwidth]{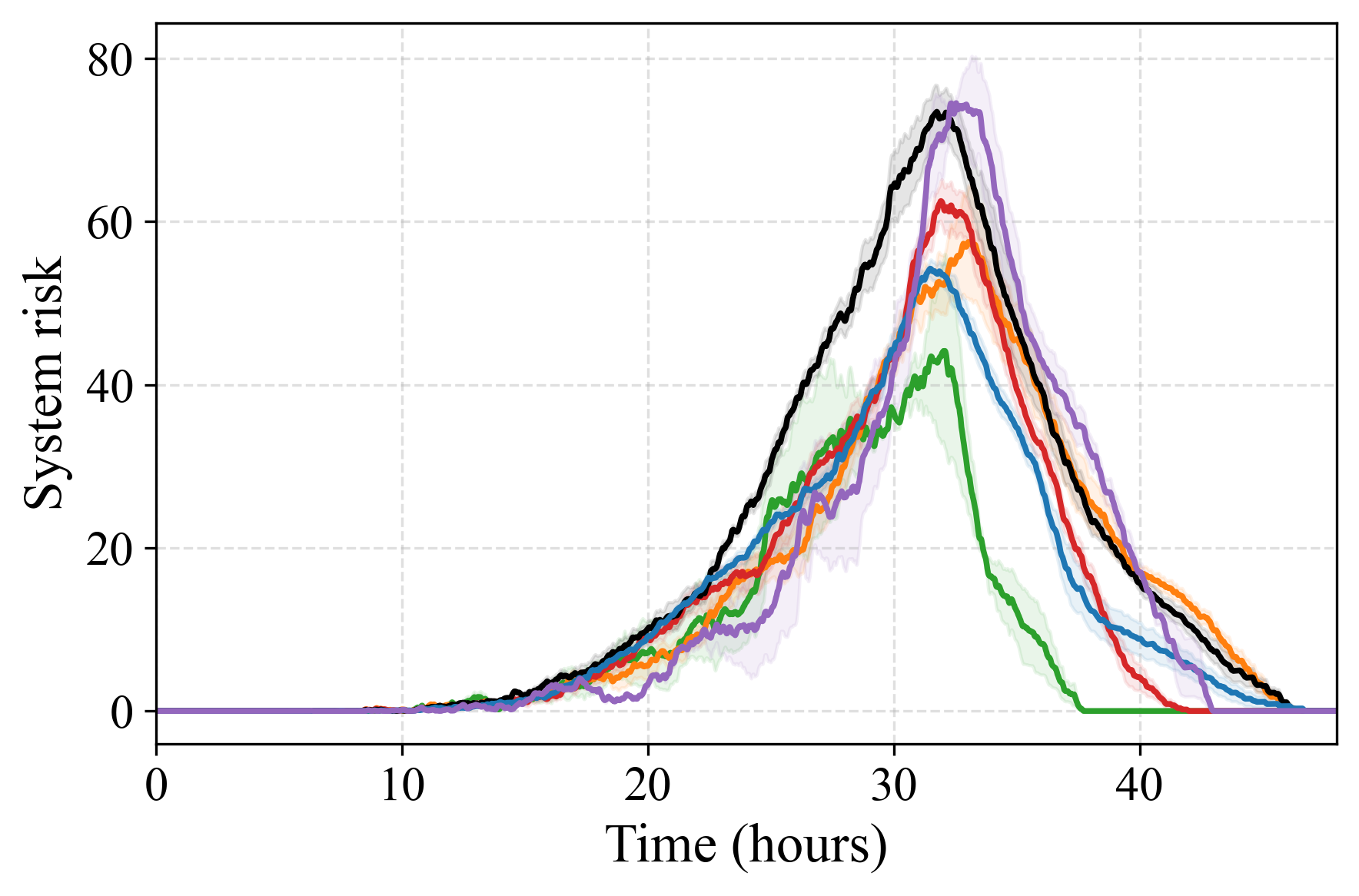}
    \label{fig:road2}
}
\hspace{-3mm}
\subfloat[Road failure - three components]{
    \includegraphics[width=0.33\textwidth]{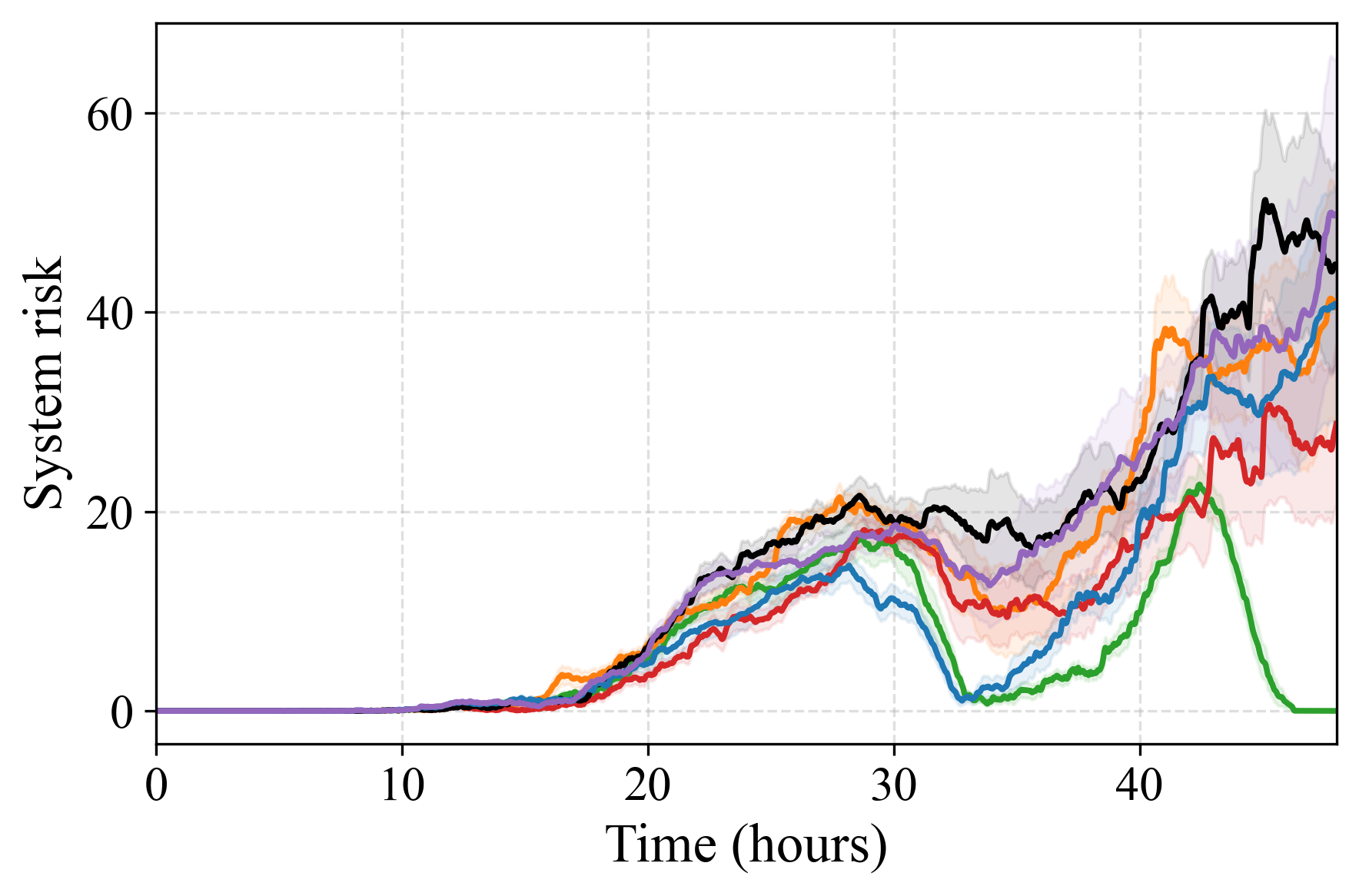}
    \label{fig:road3}
}

\caption{Temporal evolution of system risk across different scenarios: (a)-(c) evacuation demand perturbations, (d)-(f) fixed charging infrastructure failure, and (g)-(i) road link failure. System risk is measured as $\sum_{i\in\mathcal{F}} R_i(t)Q_i(t)$, where a lower value indicates better performance. Solid lines represent the mean value, and shaded bands indicate the variability across different seeds.}
\label{fig:OP_all_scenarios}
\end{figure*}

Under fixed charging infrastructure failures, ARMD achieves the best ARE and PSRE across all three failure probabilities.
To further examine how risk evolves over the evacuation process, Figure~\ref{fig:OP_all_scenarios}(d)--(f) presents the temporal evolution of system risk under different failure probabilities. As the failure probability increases from 15\% to 45\%, system risk rises more rapidly, reaches higher peaks, and remains elevated for longer durations, indicating that FCS outages amplify system vulnerability during evacuation. Across the three failure levels, ARMD consistently shows stronger peak-risk suppression than the competing methods. This advantage is clearer under the 30\% and 45\% failure scenarios, where ARMD not only reduces the risk peak but also enables efficient post-peak risk dissipation. However, under the 45\% failure scenario, the late-stage system risk remains substantial for all methods. This is likely because the effective FCS charging supply is severely constrained, and improved MCT deployment alone is insufficient to relieve queue accumulation.

\begin{figure}[!t]
\centering
\subfloat[EV travel distance]{
    \includegraphics[width=0.4\textwidth]{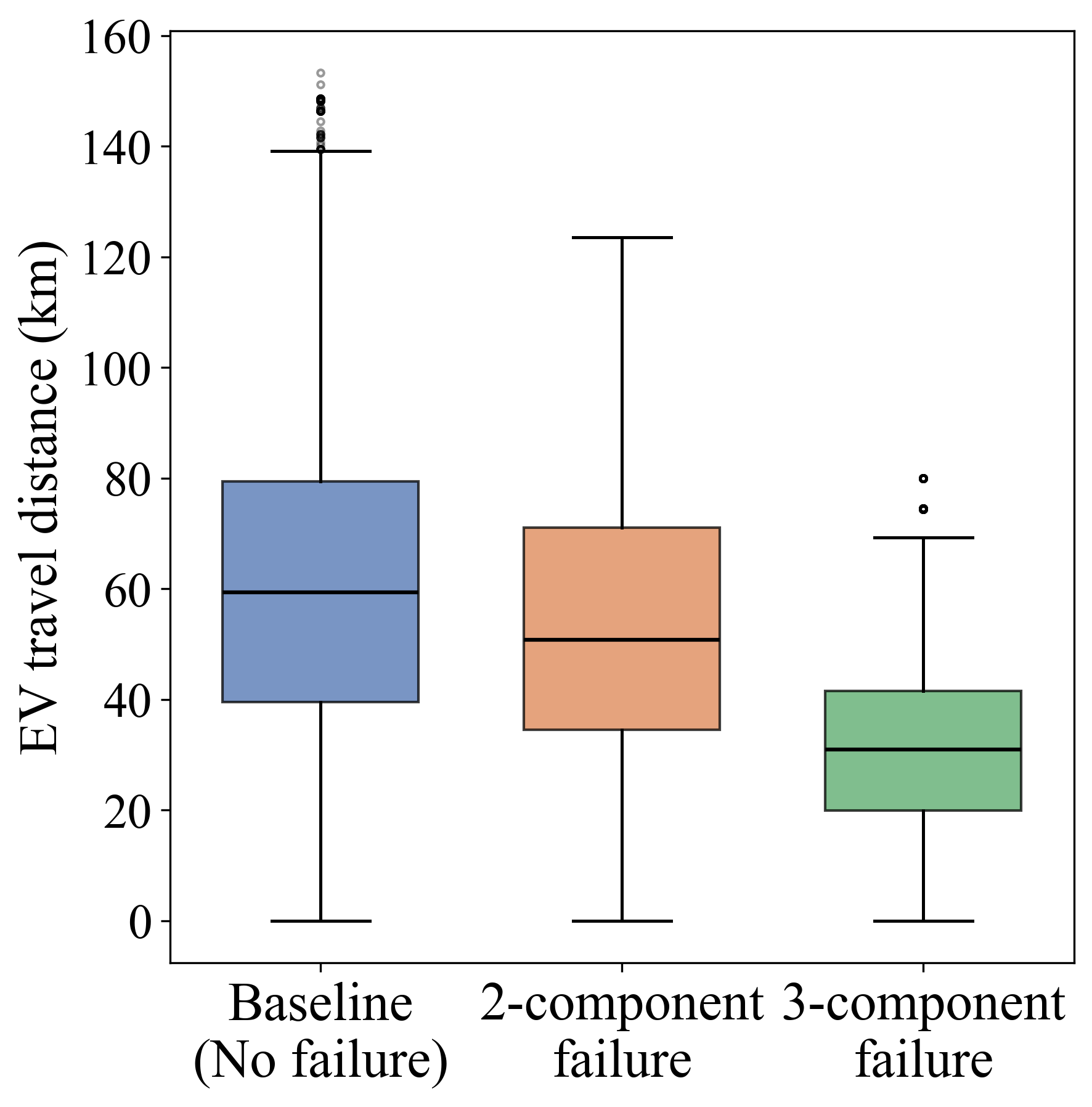}
    \label{fig:road_failure_distance}
}
\hspace{8mm}
\subfloat[Charging demand]{
    \includegraphics[width=0.4\textwidth]{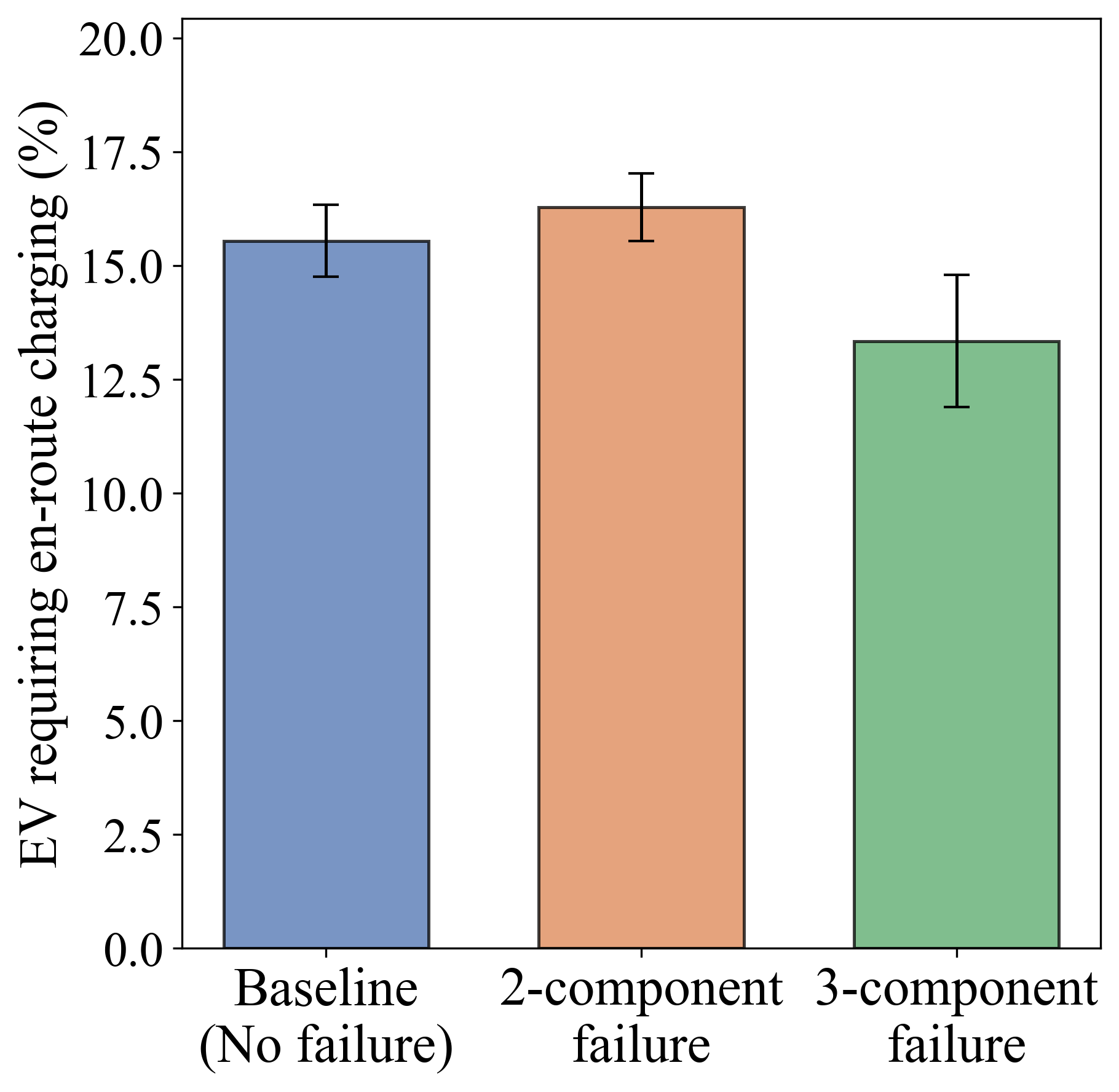}
    \label{fig:road_failure_demand}
}
\caption{Effects of road network fragmentation on EV travel distance and charging demand.}
\label{fig:road_failure_mechanism}
\end{figure}

For road link failure scenarios, ARMD obtains the lowest ARE in all three cases, suggesting that the proposed framework can maintain effective MCT deployment even under disrupted transportation networks. However, there is a seemingly counterintuitive pattern: the risks in the three-component case are lower than those in the base scenario and two-component case. This does not imply that more severe road disruption improves the overall evacuation outcome. 
Instead, it arises from changes in EV evacuees' route choices under network fragmentation. Some evacuees are rerouted to alternative destinations within their connected components, which may shorten travel distances and reduce the need for en-route charging, as illustrated in Figure~\ref{fig:road_failure_mechanism}. In addition, some evacuees may leave the simulation when no reachable safe destination exists. Both effects reduce the charging demand imposed on FCSs, thereby lowering the measured charging-system risk.

The ablation results show that both online actor fine-tuning and online route updating contribute to the overall performance of ARMD. 
In most demand-perturbation and fixed charging infrastructure failure scenarios, ARMD-NR frequently achieves the second-best results and outperforms ARMD-NF, suggesting that online actor fine-tuning plays an important role in adapting MCT deployments to changing charging demand and charger availability. In contrast, under road link failure scenarios, the advantage of ARMD-NR becomes less pronounced. This indicates that when transport network conditions are degraded, online route updating becomes more important.

\subsubsection{Comparison with Offline and Rolling-Horizon Optimization Benchmarks}
As shown in Tables~\ref{tab:performance_demand} and~\ref{tab:performance_failure}, RH-MIP generally outperforms OF-MIP, demonstrating the value of online re-optimization. However, the proposed ARMD achieves stronger performance than RH-MIP in most scenarios, especially under higher evacuation participation and more severe infrastructure failures.
The temporal evolution of system risk in Figure~\ref{fig:OP_all_scenarios} further shows that both RH-MIP and ARMD exhibit advantages over OF-MIP in suppressing risk peaks and accelerating post-peak risk dissipation, while ARMD achieves more effective peak control and faster recovery.

A possible explanation for the performance gap is that OF-MIP and RH-MIP both depend on forecasted EV arrivals, although to different extents. OF-MIP is most vulnerable because its deployment schedule is fixed over the entire evacuation horizon, so forecast errors can accumulate and propagate into middle- and late-stage deployment mismatches, leading to higher risk peaks and delayed risk dissipation. RH-MIP partially mitigates this issue through rolling re-optimization and therefore performs well in relatively mild scenarios, such as flatter departure patterns and moderate infrastructure failures. Nevertheless, RH-MIP still relies on forecast profiles over a finite planning horizon to evaluate future deployment decisions. When forecast errors become large or spatially uneven, such as under higher evacuation participation or more severe infrastructure failures, the resulting mismatch can lead to insufficient or delayed MCT support at critical FCSs, producing higher risk peaks and slower recovery. In contrast, ARMD learns risk-aware state-action mapping from stochastic evacuation trajectories, allowing its allocation decisions to adapt to realized system states without relying solely on explicit arrival forecasts during deployment.

\begin{figure}[!t]
\centering
\includegraphics[width=0.8\textwidth]{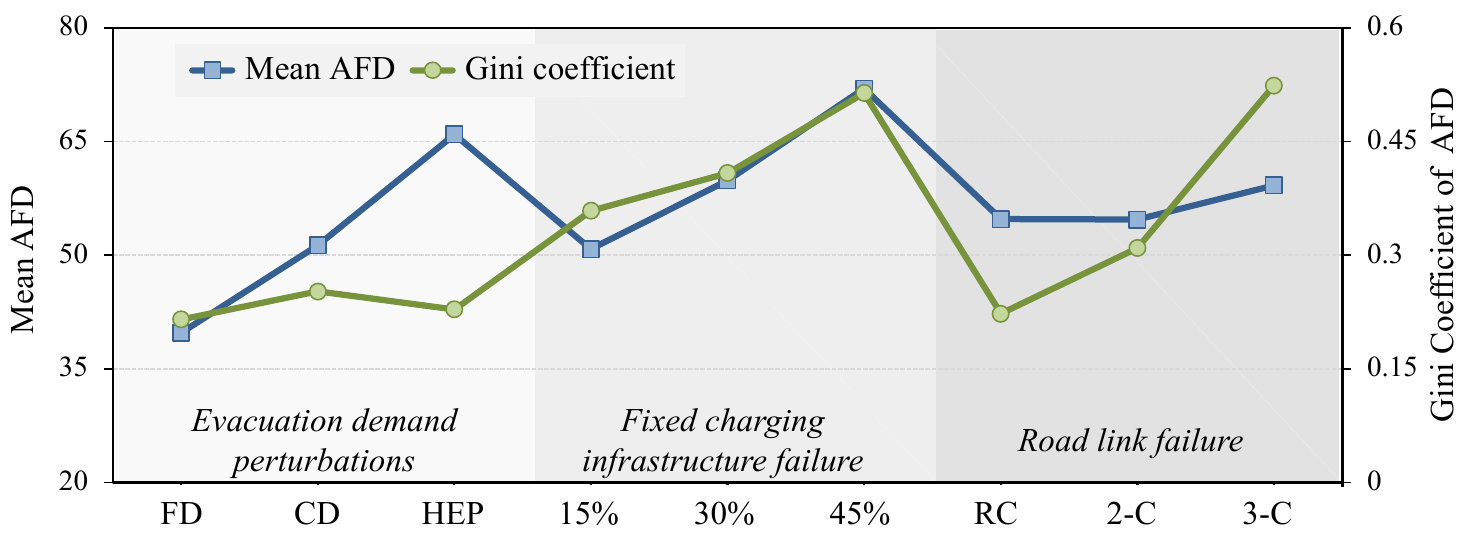}
\caption{Mean AFD and Gini coefficient of AFD under different scenarios. \textit{Note:} FD = flatter departure; CD = concentrated departure; HEP = higher evacuation participation; 15\%, 30\%, and 45\% denote station failure probabilities; RC = reduced capacity; 2-C = two components; 3-C = three components.}
\label{fig:afd}
\end{figure}

To further analyze this explanation, we introduce the arrival forecast deviation (AFD) for each FCS, defined as
\begin{equation}
\mathrm{AFD}_i=
\frac{1}{H}
\sum_{h\in\mathcal{H}}
\left|
A_{ih}^{\mathrm{real}}-A_{ih}^{\mathrm{pred}}
\right|, 
\label{eq:afd}
\end{equation}
where $h \in \mathcal{H}=\{1,\dots,H\}$ indexes the time intervals over the MCT deployment period, and $A_{ih}^{\mathrm{real}}$ and $A_{ih}^{\mathrm{pred}}$ denote the realized and predicted EV arrivals at FCS $i$ during interval $h$, respectively. AFD measures the mismatch between realized and forecasted arrivals. We compute the mean and Gini coefficient of AFD for each scenario, which capture the overall magnitude and the spatial unevenness of forecast errors. As shown in Figure~\ref{fig:afd}, higher evacuation participation and more severe failure scenarios generally exhibit larger mean AFD and higher AFD Gini coefficients. This indicates that realized arrivals deviate more substantially from the forecast profiles and that such deviations are more unevenly distributed across FCSs. These observations are consistent with the temporal risk patterns: when forecast errors become larger and more spatially concentrated, OF-MIL and RH-MIP are more likely to experience higher risk peaks and slower risk dissipation, while ARMD maintains more robust risk control.

\begin{table*}[htbp]
\centering
\footnotesize
\caption{Comparison of MCT reallocation travel time under experimental scenarios. Bold indicates a travel time reduction greater than 2 min for ARMD relative to ARMD-NR.}
\label{tab:moves_failure}
\begin{tabular}{l cc cc cc}
\toprule
& \multicolumn{6}{c}{\textit{Evacuation demand perturbations}} \\
\cmidrule(){2-7}
& \multicolumn{2}{c}{Higher evacuation participation} 
& \multicolumn{2}{c}{Concentrated departure} 
& \multicolumn{2}{c}{Flatter departure} \\
\cmidrule(lr){2-3}\cmidrule(lr){4-5}\cmidrule(lr){6-7}
& Long moves & Short moves & Long moves & Short moves & Long moves & Short moves \\
\midrule
ARMD-NR 
& $176.2\pm44.8$ & $27.2\pm6.3$  
& $119.9\pm11.6$ & $22.8\pm4.7$  
& $115.2\pm12.3$ & $26.8\pm6.1$ \\

ARMD    
& $\bm{151.5\pm46.8}$ & $\bm{22.9\pm4.7}$ 
& $118.8\pm12.0$ & $21.4\pm5.3$ 
& $\bm{112.8\pm19.8}$ & $\bm{22.4\pm3.7}$  \\
\midrule

& \multicolumn{6}{c}{\textit{Fixed charging infrastructure failure scenarios}} \\
\cmidrule(lr){2-7}
& \multicolumn{2}{c}{Failure probability = 15\%} 
& \multicolumn{2}{c}{Failure probability = 30\%} 
& \multicolumn{2}{c}{Failure probability = 45\%} \\
\cmidrule(lr){2-3}\cmidrule(lr){4-5}\cmidrule(lr){6-7}
& Long moves & Short moves & Long moves & Short moves & Long moves & Short moves \\
\midrule
ARMD-NR 
& $129.3\pm22.3$ & $25.2\pm4.3$ 
& $119.9\pm11.6$ & $22.8\pm4.7$
& $95.0\pm14.4$ & $28.6\pm2.4$ \\

ARMD    
& $\bm{120.9\pm27.6}$ & $24.7\pm1.4$ 
& $\bm{99.5\pm18.6}$ & $24.6\pm2.6$ 
& $\bm{89.2\pm4.5}$ & $26.6\pm3.6$ \\
\midrule

& \multicolumn{6}{c}{\textit{Road link failure scenarios}} \\
\cmidrule(lr){2-7}
& \multicolumn{2}{c}{Reduced capacity} & \multicolumn{2}{c}{Two components} & \multicolumn{2}{c}{Three components} \\
\cmidrule(lr){2-3}\cmidrule(lr){4-5}\cmidrule(lr){6-7}
& Long moves & Short moves & Long moves & Short moves & Long moves & Short moves \\
\midrule
ARMD-NR 
& $142.5\pm5.8$ & $23.6\pm1.7$ & $117.7\pm18.1$ & $24.0\pm2.7$ 
&  $110.7\pm22.9$ & $22.4\pm4.1$ \\

ARMD    
& \bm{$138.0\pm4.4$} & $22.6\pm3.5$ & $\bm{104.7\pm9.4}$ & $22.1\pm5.1$ 
& \bm{$97.3\pm20.8$} & $22.5\pm3.5$ \\
\bottomrule
\end{tabular}
\end{table*}

\subsubsection{Online Routing Effectiveness}
Table~\ref{tab:moves_failure} shows the average travel time of MCT reallocation trips under ARMD and ARMD-NR. 
The results indicate that the benefit of online routing is more pronounced for long reallocation moves. By contrast, its effect on short moves is generally weaker, as route options are more limited and travel times are less sensitive to dynamic re-planning over short distances. 

Across different scenarios, the advantage of online routing is most evident under the higher evacuation participation scenarios. A relatively large travel time reduction can also be observed under road link failure and charging infrastructure failure scenarios. One possible explanation is that these scenarios make traffic conditions more uneven and create larger travel-time differences across feasible routes. Higher evacuation participation raises the overall traffic demand and makes congestion more likely. Charging infrastructure failures do not directly create traffic congestion, but they can shift charging demand toward the remaining operational stations, thereby worsening traffic conditions around these stations and along their access routes. Similarly, road link failures reduce the set of feasible paths and compress traffic into fewer corridors or connected components, which increases the likelihood of localized congestion. As a result, online routing becomes more beneficial because it enables MCTs to react to these dynamically concentrated traffic conditions and avoid potentially congested links.

\subsubsection{Sensitivity Analysis on MCT Fleet Size}
\begin{figure*}[h]
\centering
\subfloat[ARE]{%
    \includegraphics[width=0.33\textwidth]{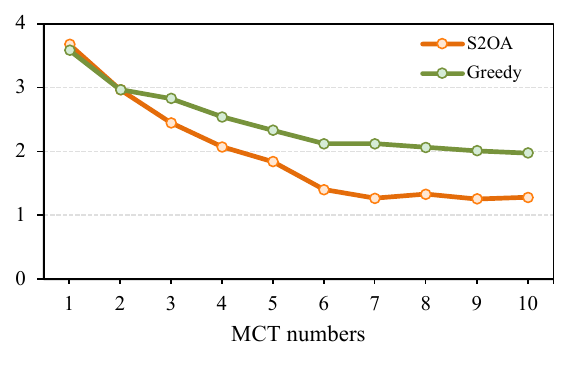}%
    \label{fig:mct_sub_a}}
\hspace{-1mm}%
\subfloat[PSRE]{%
    \includegraphics[width=0.33\textwidth]{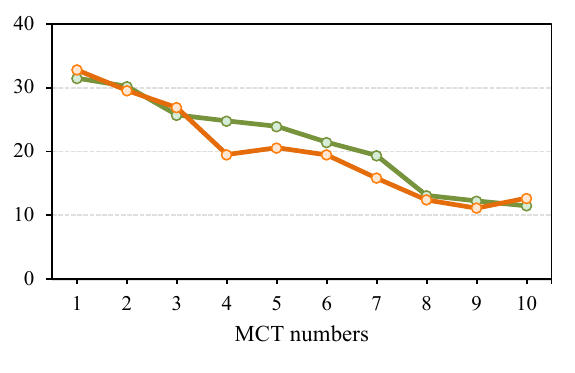}%
    \label{fig:mct_sub_b}}
\hspace{-1mm}%
\subfloat[ASRE-L]{%
    \includegraphics[width=0.33\textwidth]{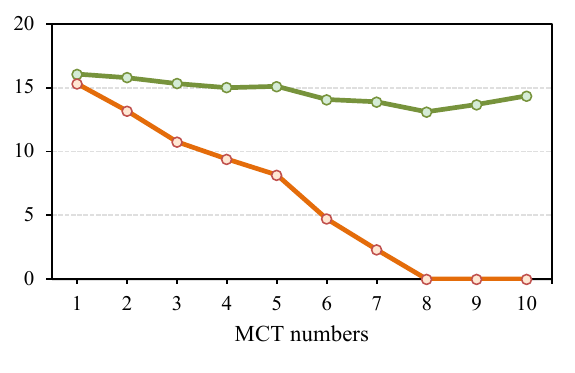}%
    \label{fig:mct_sub_c}}

\caption{Sensitivity analysis of MCT fleet size under ARMD and Greedy.}
\label{fig:mct_supply}
\end{figure*}

To examine how the number of MCTs affects evacuation performance, we conducted a sensitivity analysis by varying the number of deployed MCTs. In this experiment, 
the pre-trained policy with six MCTs is directly deployed to scenarios with different fleet sizes without retraining.  Figure~\ref{fig:mct_supply} compares ARMD with the Greedy baseline under varying MCT fleet sizes. Note that CA is not included in this analysis, because pretrained centralized policy cannot be directly transferred to scenarios with different fleet sizes.

Overall, increasing the MCT fleet size reduces evacuation risk, but the marginal benefit gradually diminishes at higher supply levels. In the simulated scenarios, 
the reduction in risk metrics is more substantial when the fleet size increases from a low level to a moderate level, with noticeable improvements observed up to around seven MCTs. Beyond this level, increasing the fleet size from 8 to 10 MCTs yields smaller incremental improvements. This suggests that once mobile charging supply becomes sufficient for the major charging bottlenecks in the simulated scenarios, additional MCTs may become underutilized, leading to diminishing marginal benefits. However, the value of seven MCTs is a scenario-dependent observation rather than a universal fleet-size threshold. 

When the fleet size of MCTs is very small, ARMD and Greedy exhibit similar performance. As the the number of MCTs increases, the superiority of ARMD becomes more apparent, particularly in ARE and ASRE-L. 
Figure~\ref{fig:mct_supply}\subref{fig:mct_sub_c} also shows the effectiveness of ARMD in mitigating late-stage risk exposure by driving ASRE-L close to zero when the fleet size becomes sufficiently large. 
This suggests that the proposed decentralized policy retains good transferability and remains effective even when the available MCT supply changes after deployment.

%% file: implication.tex
\section{Real-world Implications of the Proposed ARMD Framework}
\emph{Application to evacuations and other large-scale events:}
The proposed ARMD framework provides an online decision-support tool for emergency management agencies to deploy MCTs under evolving evacuation conditions. In practice, agencies can tailor the simulator using local transportation networks, evacuation demands, and MCT fleet configurations, and pre-train the allocation policy for region-specific operating conditions. During an actual evacuation, the pre-trained policy can be updated using real-time charging and traffic information to support adaptive MCT deployment following the ARMD design. Although this study is motivated by evacuation during natural disasters, the framework can also support other scenarios in which EV charging demand experiences a sudden surge beyond routine patterns, such as holiday peak-travel periods.

Beyond mobile charging support, this adaptive framework can be extended to other online resource deployment problems. For example, during mega events, police patrol units or emergency response teams may need to be dynamically deployed in response to time-varying crowd density, traffic incident reports, and public safety complaints. Similar to MCT deployment, these problems require adaptive deployment of limited mobile resources responding to rapidly evolving operational conditions. By replacing evacuees' risk exposure with event-specific risk or service-demand indicators, the ARMD framework can provide a generalizable structure for adaptive patrol or emergency resource allocation under uncertainty.

\emph{Integration with disaster digital twin:}
The proposed framework can also be integrated into a disaster digital twin (DDT) \cite{ghaffarian2025rethinking} for resilient emergency management for complex disaster conditions. Urban infrastructure systems are highly interconnected. Transportation networks, power supply networks, communication networks, and social communities are all critical entities at risk during disasters \cite{lagap2024digital}. Therefore, a DDT needs to fuse heterogeneous but interdependent system information and maintain an updated representation of the evolving disaster environment. In this study, time-varying risk exposure is simplified as an exponential function. Within a DDT, however, multi-source observations can be integrated with disaster evolution models to generate higher-fidelity risk estimates. These estimates can update the simplified risk representation in the state and reward functions of MAPPO, enabling MCT deployment decisions to better reflect the evolving hazard environment. In addition, charging efficiency may be affected by power-grid stress and voltage instability in practice. By incorporating real-time grid information, the state representation can be extended to include dynamic charging-efficiency attributes at each FCS. This would allow the MAPPO-based allocation policy to account for potential reductions in charging capacity caused by grid conditions.

The framework can also be extended beyond evacuees’ charging demand to critical facilities with urgent power needs, such as shelters, emergency operation centers, medical sites, and communities. This extension can be implemented by incorporating facility locations into the transportation network and adding facility-specific power demand and deprivation-cost indicators to the system state. The deployment objective can then be modified to balance EV charging-queue risk with critical-facility power support. In this way, integration with DDTs can broaden the applicability of ARMD and strengthen emergency charging and power-support resilience under compound disaster disruptions.

%% file: conclusion.tex
\section{Conclusion}
In this paper, we propose ARMD, an online framework for MCT deployment to mitigate evacuees' risk exposure induced by charging congestion at overloaded FCSs during large-scale evacuations. 
It formulates MCT deployment as a dynamic, risk-aware decision-making problem under evolving evacuation conditions, rather than as a static placement problem. 
Unlike traditional methods, ARMD combines offline learning with online updating, enabling MCT deployment decisions to adapt to stochastic evacuation environments.
It divides MCT deployment into two coupled decision problems: allocation and routing. For MCT allocation, a MAPPO-based policy is learned offline and fine-tuned online to capture the evolution of charging demands and risks. 
To support online MCT routing, the STPM is trained offline to anticipate short-term traffic dynamics. After each MCT is assigned to a target FCS, its relocation route is updated in rolling horizon using the short-term travel time forecasts of STPM. This offline-to-online design enhances the operational adaptability and resilience of evacuation charging support under uncertain and evolving conditions. The experiments conducted on an evacuation simulator built from Hillsborough County, Florida, demonstrate the effectiveness of the proposed framework. Across a range of evacuation demand perturbation and infrastructure disruption scenarios, ARMD consistently maintains relatively low risk exposure. These results indicate that the proposed framework can achieve efficient MCT deployments and enhance the resilience of emergency charging support under uncertain hurricane evacuation conditions.

There are several limitations that should be acknowledged. First, the present study is developed and evaluated in a simulated evacuation environment, mainly due to the limited availability of real-world traffic and charging-demand data during large-scale evacuations. Although the simulator captures hurricane-risk dynamics and traffic conditions, future work should further validate and recalibrate the proposed framework using real-world data when such data become available.  
Second, the current framework assumes fixed service duration for each MCT after arriving at an FCS. This assumption simplifies the operational process but may limit the flexibility of MCT responses to heterogeneous and time-varying station conditions. Future research can relax this assumption by introducing an adaptive service-duration mechanism into the MAPPO policy. Specifically, after an MCT arrives at an FCS, its service duration can be continuously reassessed according to the observed charging demand and remaining onboard energy, rather than being fixed in advance. 

%% file: appendix.tex
\appendix
\section{MIP Model}
\label{MILP_model}
This appendix presents the MIP model used by the OF-MIP and RH-MIP benchmarks. Let $\mathcal{P}=\{0,1,\dots,M-1\}$ denote the set of decision epochs covered in the planning horizon, where $M$ is the number of decision epochs. Each decision epoch $p\in\mathcal{P}$ starts at time $\tau_p$ and has a fixed duration $\Delta$. MCT allocation decisions are made at the beginning of each decision epoch. To obtain a tractable optimization model, the queue dynamics in FCSs in each epoch is approximated as a linear process. 

\begin{table}[h]
\caption{Notation used in the MIP model.}
\label{tab:milp_notation}
\centering
\renewcommand{\arraystretch}{1.08}
\begin{tabular}{p{0.08\linewidth} p{0.82\linewidth}}
\hline
\textbf{Symbol} & \textbf{Description} \\
\hline
\multicolumn{2}{l}{\textit{Sets and indices}}\\
$\mathcal{F}$ & set of FCSs. \\
$\mathcal{K}$ & set of MCTs. \\
$\mathcal{P}$ & set of decision epochs in the planning horizon. \\
\hline
\multicolumn{2}{l}{\textit{Parameters}} \\
$A_i^p$ & forecast EV arrivals at FCS $i$ during epoch $p$. \\
$m_i^p$ & number of available fixed chargers at FCS $i$ during epoch $p$. \\
$Q_i^0$ & initial queue length at FCS $i$ at epoch $0$. \\
$\Delta$ & duration of each decision epoch. \\
$\mu^{\mathrm{FCS}}$ & service rate of one fixed charger. \\
$\mu^{\mathrm{MCT}}$ & service rate of one MCT. \\
$L_{ki}^{0}$ & travel time from the location of MCT $k$ to FCS $i$ at epoch $0$. \\
$L_{kji}^{p}$ & forecast relocation travel time for MCT $k$ from FCS $j$ to FCS $i$ at epoch $p$, $p\ge 1$. \\
$u_k^0$ & initial charging capability of MCT $k$. \\
\hline
\multicolumn{2}{l}{\textit{Decision variables}} \\
$x_{ki}^p$ & binary variable equal to 1 if MCT $k$ is assigned to FCS $i$ at epoch $p$. \\
$z_{kji}^p$ & binary variable equal to 1 if MCT $k$ moves from FCS $j$ in epoch $p-1$ to FCS $i$ in epoch $p$. \\
$U_{ki}^p$ & service amount provided by MCT $k$ at FCS $i$ during epoch $p$. \\
$S_i^p$ & service amount provided by fixed chargers at FCS $i$ during epoch $p$. \\
$Q_i^p$ & queue length at FCS $i$ at the beginning of epoch $p$. \\
\hline
\end{tabular}
\end{table}

The notations used for the MIP are summarized in Table~\ref{tab:milp_notation}. The objective is to minimize the cumulative risk exposure of EV evacuees queuing at FCSs over the planning horizon:
\begin{equation}
\min 
\sum_{p\in\mathcal{P}}
\sum_{i\in\mathcal{F}}
\int_{\tau_p}^{\tau_{p+1}} R_i(t)Q_i(t)\,\mathrm{d}t .
\label{eq:mip_obj}
\end{equation}

Within each decision epoch $p$, the queue trajectory is approximated by linear interpolation:
\begin{equation}
Q_i(t)
=
Q_i^p+
\frac{t-\tau_p}{\Delta}
\left(Q_i^{p+1}-Q_i^p\right),
\qquad t\in[\tau_p,\tau_{p+1}].
\label{eq:mip_queue_linear}
\end{equation}

Since $R_i(t)$ is exponential as defined in Section~\ref{sim_setup} and $Q_i(t)$ is linear, the integral in Eq.~\eqref{eq:mip_obj} admits a closed-form expression. The constraints are given as follows:
\begin{equation}
\sum_{i\in\mathcal{F}}x_{ki}^p=1
\qquad \forall k, p
\label{eq:mip_assign}
\end{equation}
\begin{equation}
z_{kji}^p \le x_{kj}^{p-1}
\qquad \forall k, j, i,\; p\in\mathcal{P}\setminus\{0\}
\label{eq:mip_trans1}
\end{equation}
\begin{equation}
z_{kji}^p \le x_{ki}^{p}
\qquad \forall k, j, i,\; p\in\mathcal{P}\setminus\{0\}
\label{eq:mip_trans2}
\end{equation}
\begin{equation}
z_{kji}^p \ge x_{kj}^{p-1}+x_{ki}^{p}-1
\qquad \forall k, j, i,\; p\in\mathcal{P}\setminus\{0\}
\label{eq:mip_trans3}
\end{equation}
\begin{equation}
\ell_{ki}^{0}=\max\{0,\Delta-L_{ki}^{0}\}
\qquad \forall k, i
\label{eq:effective_service_time_initial}
\end{equation}
\begin{equation}
0\le U_{ki}^{0}\le
\mu^{\mathrm{MCT}}\ell_{ki}^{0}x_{ki}^{0}
\qquad \forall k, i
\label{eq:mip_mct_service_initial}
\end{equation}
\begin{equation}
\ell_{kji}^{p}=\max\{0,\Delta-L_{kji}^{p}\}
\qquad \forall k,j,i,\; p\in\mathcal{P}\setminus\{0\}
\label{eq:effective_service_time}
\end{equation}
\begin{equation}
0\le U_{ki}^{p}\le
\mu^{\mathrm{MCT}}
\sum_{j\in\mathcal{F}}
\ell_{kji}^{p}
\, z_{kji}^{p}
\qquad \forall k, i,\; p\in\mathcal{P}\setminus\{0\}
\label{eq:mip_mct_service_future}
\end{equation}
\begin{equation}
0\le S_i^{p}
\le
\mu^{\mathrm{FCS}}m_i^{p}\Delta
\qquad \forall i, p
\label{eq:mip_fixed}
\end{equation}
\begin{equation}
S_i^{p}+\sum_{k\in\mathcal{K}}U_{ki}^{p}
\le Q_i^{p}+A_i^{p}
\qquad \forall i, p
\label{eq:mip_service}
\end{equation}
\begin{equation}
Q_i^{p+1}
=
Q_i^{p}+A_i^{p}-S_i^{p}
-\sum_{k\in\mathcal{K}}U_{ki}^{p}
\qquad \forall i,\; p=0,\dots,M-2
\label{eq:mip_queue}
\end{equation}
\begin{equation}
\sum_{p\in\mathcal{P}}
\sum_{i\in\mathcal{F}}
U_{ki}^{p}
\le u_k^0
\qquad \forall k
\label{eq:mip_cap}
\end{equation}
\begin{equation}
x_{ki}^{p}\in\{0,1\}
\qquad \forall k, i, p
\label{eq:mip_domain_x}
\end{equation}
\begin{equation}
z_{kji}^{p}\in\{0,1\}
\qquad \forall k, j, i,\; p\in\mathcal{P}\setminus\{0\},
\label{eq:mip_domain_z}
\end{equation}
\begin{equation}
U_{ki}^{p}\ge0,\quad S_i^{p}\ge0,\quad Q_i^{p}\ge0
\qquad \forall k, i, p
\label{eq:mip_domain_cont}
\end{equation}

Constraints~\eqref{eq:mip_assign} guarantees that each MCT is assigned to one FCS in each decision epoch. 
Constraints~\eqref{eq:mip_trans1}--\eqref{eq:mip_trans3} capture the relocation of MCTs between two consecutive decision epochs. Specifically, $z_{kji}^p$ takes value 1 if and only if MCT $k$ is assigned to FCS $j$ in epoch $p-1$ and to FCS $i$ in epoch $p$.
Constraints~\eqref{eq:effective_service_time_initial}--\eqref{eq:mip_mct_service_future} restrict the service amount provided by each MCT based on its effective service time after relocation.
Constraint~\eqref{eq:mip_fixed} limits the service amount provided by fixed chargers at FCSs according to its available chargers.
Constraint~\eqref{eq:mip_service} ensures that the total service provided at each FCS does not exceed the charging demand.
Constraint~\eqref{eq:mip_queue} describes the queue evolution across decision epochs. 
Constraint~\eqref{eq:mip_cap} ensures that the total service provided by each MCT does not exceed its charging capability. 
Finally, constraints~\eqref{eq:mip_domain_x}--\eqref{eq:mip_domain_cont} define the domains of the decision variables.

In this study, each decision epoch spans $2.5$ hours, which is designed to cover the $2$-hour MCT service cycle and the travel time required for reallocation. The forecast profiles used in the optimization benchmarks are approximated by empirical mean arrivals and travel times obtained from multiple simulation runs without MCT deployment. The optimization problems are solved using Gurobi.
For OF-MIP, the optimization horizon covers the entire evacuation period, i.e., up to the end of the $48$-hour evacuation window. The model is solved once before deployment and outputs a fixed MCT deployment plan for all decision epochs. This plan is then applied throughout the evacuation without online re-optimization. In infrastructure failure scenarios, if a pre-planned target FCS becomes unavailable or unreachable during execution, a simple feasibility repair is applied by redirecting the MCT to the closest reachable FCS. 
For RH-MIP, the planning horizon includes three decision epochs. At each decision epoch, the model outputs MCT deployment decisions over this finite horizon, but only the decisions for the current epoch are implemented. At the next decision epoch, the system state and forecast profiles are updated, and the model is re-solved.